\pdfoutput=1
\documentclass{article}

\PassOptionsToPackage{numbers,compress}{natbib}
\usepackage[eandd,final]{neurips_2026}

\usepackage[utf8]{inputenc}
\usepackage[T1]{fontenc}
\usepackage{hyperref}
\usepackage{url}
\usepackage{booktabs}
\usepackage{amsfonts}
\usepackage{amsmath}
\usepackage{nicefrac}
\usepackage{xcolor}
\usepackage{graphicx}
\usepackage{multirow}
\usepackage{placeins}

\title{VIPBench: A Human-Aligned Benchmark for Voice Identity Perception in the Age of Voice Cloning}

\author{%
  Xudong Tang \\
  Department of Computer Science \\
  Northwestern University
  \And
  Matthew Groh\thanks{Corresponding author.} \\
  Management and Organizations \\
  Kellogg School of Management, Northwestern University
}

\begin{document}

\maketitle

\begin{abstract}

Speaker-recognition benchmarks test how accurately a model recovers the speaker label of a recording, treating voice identity as a property of the person who produced it. However, these benchmarks do not evaluate how humans perceive voice identity. Whether a listener hears two voices as the same person has consequences for intellectual property rights, cybersecurity, and everyday life. In this paper, we make two contributions. First, we introduce the Voice Identity Perception benchmark (\textbf{VIPBench}), which contains 124{,}876 same-or-different judgments from 1{,}290 listeners on 9{,}800 recording pairs from 100 public figures. The pairs include real recordings, voice clones generated by a text-to-speech (TTS) system, and morphs between two speakers' voices. Second, we define four tasks that evaluate the alignment between speaker-representation models and human identity perception. The first two tasks measure how well a model's similarity score predicts the proportion of listeners who respond ``same'' and the majority response. The other two measure whether a model and listeners find the same speaker pairs similar, and whether a model's relationship to listener judgments holds equally for real and synthetic recordings. We evaluate eleven publicly available speech models. Listener judgments diverge from speaker labels most often for voice clones: in 41\% of responses, listeners judged a clone and a real recording of the same speaker to be different people. No model reaches the noise ceiling; the best model explains 61\% of the explainable variance. Models trained with speaker labels align with listeners better than self-supervised models do, yet a model with a 12\% error rate on speaker labels aligns with listeners as well as models with error rates below 2\%. VIPBench enables the evaluation of speaker models against human perception on real, cloned, and morphed speech, and our results motivate training speaker models directly on listener judgments.
\end{abstract}

\section{Introduction}
A person's voice is often treated as a credential of their identity. For decades, the core goal of automatic speaker verification and voice biometric authentication systems has been to identify speakers from their voices more accurately. These systems are now widely used to secure financial accounts, personalized smart speakers, and mobile devices~\citep{reynolds2002overview, hansen2015speaker, bai2021speaker}, and to provide evidence in forensic casework~\citep{neustein2012forensic}. Traditionally, these systems have been trained almost exclusively on speaker labels from recording metadata, which define each recording's target identity as the speaker who, in ground truth, produced it. Humans, however, do not always perceive a voice as belonging to the speaker in its metadata label. Even a subtle change to one speaker's voice, such as a slight increase in fundamental frequency, can alter the identity a listener perceives~\citep{baumann2010perceptual}, a well-documented consequence of within-speaker variability~\citep{lavan2019flexible,kanber2022highly} in voice perception research.

For recordings of real voices, training verification systems only on metadata labels is well justified. However, modern text-to-speech (TTS) systems challenge this convention. From only a short reference recording, they can generate high-fidelity voice clones that listeners cannot reliably identify as synthetic~\citep{Mai_2023,groh2024human,Barrington2025VoiceClones}. By design, a voice clone shares the metadata identity of the speaker it was generated from, since it is built to replicate that speaker's voice as closely as possible. However, just as subtle changes to a real voice can shift the identity listeners perceive, listeners may also treat a clone's slight differences from its source as evidence of a different speaker. Consider a bank's voice authentication system trained on metadata alone, which would likely accept clones as the customer, including clones listeners would hear as someone else. A perception-aligned model can flag those clones, providing a defense layer that metadata-based verification cannot.

Existing speech benchmarks rarely include human perception data. Most benchmarks use metadata labels as targets, which are either speaker labels indicating who produced the recordings~\citep{nagrani2017voxceleb,chung2018voxceleb2,voxblink2}, or real-versus-synthetic flags for cloned or synthesized speech~\citep{asvspoof2021,fakeavceleb}. The closest voice-domain work with human perception labels as targets is \textit{VoiceMorph}~\citep{shimizu2025voicemorph}, a behavioral study of self-voice recognition with 21 participants that uses AI-cloned and continuously morphed stimuli, for which no clean metadata speaker label exists. In adjacent domains~\citep{GERLACH202085, Chernyak2024perceptual, Kim2025predicting}, particularly vision~\citep{lu2023seeing, hebart2023things}, large-scale human annotation has proven valuable for evaluating how well learned representations align with human perception. In the voice domain, however, no prior work provides perception-aligned voice identity labels at the scale needed to benchmark machine voice representations.

To fill this gap in the speech domain, we introduce \textbf{VIPBench}, a dataset of 124{,}876 binary identity judgments from 1{,}290 listeners on 9{,}800 voice pairs over 100 speakers that span diverse linguistic backgrounds among English speakers in the United States. The pairs cover real recordings, voice clones generated with a state-of-the-art TTS system~\citep{cartesia_sonic}, and 8{,}100 continuously morphed voices. We also release a small held-out extension for robustness checks, in which 152 new listeners judged 60 pairs from 10 new speakers, with clones generated by a second TTS system~\citep{du2024cosyvoice2scalablestreaming}. We then define an evaluation protocol with four tasks that measure how well a model's speech representations align with human perception. We apply this protocol to eleven publicly available speech representation models trained under different paradigms, including supervised and self-supervised models and one weakly supervised model.

We present four main findings. First, listener perception diverges from metadata labels, especially for voice clones, where listeners judged a clone and a real recording of its source speaker to be different speakers in 41\% of responses, even though the metadata label of the pair is the same speaker. Second, no model we evaluated reaches the noise ceiling. The best model reaches 61\% of the ceiling and agrees with the listener majority more often than 72\% of individual listeners do. Third, representations from supervised models align better with listeners' perception than those from self-supervised learning (SSL) models, and SSL models align best at their earliest layers. Fourth, human perception and metadata labels rank the three model types in the same order, but their rankings diverge within each type. The top four supervised models on metadata labels all have equal error rates (EERs) below 2\%, yet \textit{resemblyzer}, with a metadata EER of 12\%, aligns with listener perception as well as any of them.

\section{Related work and benchmark positioning}

\textbf{Speaker-verification benchmarks.} 
Existing speaker-verification benchmarks, such as \textit{VoxCeleb1}~\citep{nagrani2017voxceleb}, \textit{VoxCeleb2}~\citep{chung2018voxceleb2}, and \textit{VoxBlink2}~\citep{voxblink2}, consist of hundreds of thousands to millions of utterances from large speaker pools, each labeled with its speaker identity from metadata (Table~\ref{tab:benchmark-positioning}). These benchmarks are widely used in speaker-verification and voice-biometrics research, where speaker embeddings are typically evaluated using the equal error rate (EER) and the minimum detection cost function (minDCF). However, they do not include human perceptual labels for natural voices, voice clones, or voice morphs. VIPBench adopts the same pair-level identity-evaluation framework as these benchmarks but replaces the metadata-derived target with a human-perception target.

\textbf{Supervised speaker representations.} 
A speaker representation is an embedding vector extracted from a single utterance. Early representations include i-vectors~\citep{Dehak2011}, which are derived from a factor-analysis model. More recent representations are produced by neural encoders such as x-vectors~\citep{Snyder2018}, ECAPA-TDNN~\citep{Desplanques_2020}, RawNet3~\citep{Jung2022}, and TitaNet~\citep{Koluguri2022}, which are trained with classification or metric-learning objectives such as the generalized end-to-end (GE2E) loss~\citep{wan2018ge2e}. Open-source toolkits such as WeSpeaker~\citep{wang2022wespeakerresearchproductionoriented} provide implementations of many of these models. These embeddings are designed to discriminate between speakers for verification tasks.

\textbf{General speech representations.} 
Self-supervised models (wav2vec~2.0~\citep{baevski2020wav2vec2}, HuBERT~\citep{hsu2021hubert}, WavLM~\citep{chen2022wavlm}, and XLS-R~\citep{babu2022xlsr}) and the weakly supervised Whisper model~\citep{radford2023whisper} are not optimized with speaker labels. They learn general-purpose speech representations. Different layers of these models encode different kinds of information and are therefore suited to different downstream tasks~\citep{yang2021superb}, with speaker information concentrated in the early-to-middle layers. We select the best layer for each model using nested cross-validation with speaker-disjoint folds.

\textbf{Deepfake and spoofing benchmarks.}
Deepfake corpora that include synthetic speech, such as ASVspoof 2021~\citep{asvspoof2021}, WaveFake~\citep{frank2021wavefake}, FakeAVCeleb~\citep{fakeavceleb}, DeepSpeak~\citep{barrington2024deepspeak}, SpeechFake~\citep{huang2025speechfake}, SCDF~\citep{stanek2025scdf}, and IndieFake~\citep{kumar2025indiefake}, are designed for training automatic deepfake detectors, where each recording is labeled as real or synthetic (Table~\ref{tab:benchmark-positioning}). These corpora do not contain listener judgments and are rarely designed around paired comparisons. VIPBench also includes AI voice clones, and its celebrity speakers are stratified across sociophonetic groups. However, VIPBench targets human perception of who is speaking rather than real-versus-synthetic detection. This distinction matters because a voice clone can be indistinguishable from a real recording yet still fail to sound like the target speaker.

\textbf{Perceptual audio benchmarks.} 
The closest perceptual audio benchmark to VIPBench is the Voice Conversion Challenge (VCC) series~\citep{lorenzotrueba2018vcc,zhao2020vcc,huang2023svcc}. These challenges evaluate voice-conversion systems using crowdsourced same-speaker similarity judgments, the same perceptual target as VIPBench. VCC~2018 collected 7{,}716 four-point similarity ratings from 267 listeners covering 12 speakers; VCC~2020 collected 29{,}760 four-point similarity ratings from 274 native and non-native listeners covering 8 speakers; and the Singing Voice Conversion Challenge (SVCC)~2023 extended the series to singing voice (Table~\ref{tab:benchmark-positioning}). VIPBench goes beyond VCC in both scale (124{,}876 judgments from 1{,}290 listeners on 9{,}800 voice pairs from 100 sociophonetically stratified speakers) and coverage (900 pairs of real voices, 800 pairs of state-of-the-art voice clones, and 8{,}100 pairs of continuously morphed voices). Other perceptual audio benchmarks focus on different perceptual targets. CDPAM~\citep{manocha2021cdpam} contains approximately 30{,}000 triplet judgments of signal degradations for evaluating audio-quality metrics, and \citet{Chernyak2024perceptual} compare self-supervised speech representations with intelligibility judgments of spoken words.

\textbf{Human voice identity perception.}
How listeners perceive voice identity has been a central question in cognitive science and linguistics~\citep{belin2011understanding,kreiman2011foundations}. Perceptual studies show that listeners are biased in perceiving two recordings of the same speaker as coming from two different speakers~\citep{lavan2019flexible,Lavan2019}, and that this bias is reduced when listeners are familiar with the speaker~\citep{vanlancker1987voice,stevenage2018drawing}. Recognizing a familiar speaker draws on long-term memory~\citep{brown1979memory}, whereas matching unfamiliar voices relies on comparing bundles of acoustic features and is considerably less accurate~\citep{kreiman2011foundations, Lavan2019}. Previous work also shows that voice identity is coded relative to a prototype, where a voice is represented by its distance from an internal average of voices, which is shaped by the listener's experience~\citep{latinus2011anti,LATINUS20131075,lavan2019listeners}. Speaker representations from models likewise place each voice in a high-dimensional space, where the cosine similarity between two representations measures how similar the two voices are. VIPBench therefore tests whether this similarity aligns with listener judgments.

\begin{table}[t]
\centering
\small
\caption{The table compares VIPBench with related benchmarks. The speaker-verification and deepfake benchmarks have metadata labels and no human judgments. CDPAM has human judgments of audio quality, and Fake2M and THINGS-data are image benchmarks with human judgments from an adjacent field. The Voice Conversion Challenges (VCC) and VoiceMorph also have human identity judgments, with fewer judgments and listeners than VIPBench, and only VoiceMorph has morphs. VIPBench is the first voice benchmark with human identity judgments on real recordings, AI voice clones, and voice morphs at a scale of more than 100{,}000 judgments. The columns are the size of the source corpus (\emph{\# Speakers}, \emph{\# Stimuli}), the human annotations (\emph{Human judgments}, \emph{\# Participants}), the stimulus content (\emph{AI-gen}, \emph{Morphs}), and the evaluation \emph{Target}. A check mark means that the benchmark has what the column names, and a cross means that it does not. A dash means that the entry does not apply or was not publicly reported. The citations are in Sections~1 and~2.}
\label{tab:benchmark-positioning}
\setlength{\tabcolsep}{4pt}
\resizebox{\textwidth}{!}{%
\begin{tabular}{llcccccccl}
\toprule
Benchmark & Domain & \# Speakers & \# Stimuli & \textbf{Human judgments} & \# Participants & AI-gen & Morphs & Target \\
\midrule
VoxCeleb 1+2 & Voice & 7{,}363 & 1.28M & $\times$ & $\times$ & $\times$ & $\times$ & Speaker ID \\
VoxBlink2 & Voice & 111{,}284 & 9.9M & $\times$ & $\times$ & $\times$ & $\times$ & Speaker ID \\
ASVspoof 2021 & Voice & -- & 1.4M & $\times$ & $\times$ & $\checkmark$ & $\times$ & Real/synth \\
WaveFake & Voice & 2 & 105K & $\times$ & $\times$ & $\checkmark$ & $\times$ & Real/synth \\
FakeAVCeleb & Voice+Vid & 500 & 20K & $\times$ & $\times$ & $\checkmark$ & $\times$ & Real/synth \\
DeepSpeak & Voice+Vid & 500 & 13{,}025 & $\times$ & $\times$ & $\checkmark$ & $\times$ & Real/synth \\
SpeechFake & Voice & 720 & 3.3M & $\times$ & $\times$ & $\checkmark$ & $\times$ & Real/synth \\
SCDF & Voice & 50 & 237K & $\times$ & $\times$ & $\checkmark$ & $\times$ & Real/synth \\
IndieFake & Voice & 50 & 19{,}560 & $\times$ & $\times$ & $\checkmark$ & $\times$ & Real/synth \\
CDPAM & Audio & -- & -- & 85K & -- & $\times$ & -- & Audio quality \\
Fake2M/HPBench & Image & -- & 2M$^+$ & 5K & 50 & $\checkmark$ & -- & Real/synth \\
THINGS-data & Image & -- & 26{,}107 & 4.70M & 12{,}340 & $\times$ & -- & Visual sim \\
VCC 2018 & Voice & 12 & -- & 7{,}716 & 267 & $\checkmark$ & $\times$ & Identity perception \\
VCC 2020 & Voice & 8 & -- & 29{,}760 & 274 & $\checkmark$ & $\times$ & Identity perception \\
VoiceMorph & Voice & 42 & 2{,}163 & 2{,}751 & 21 & $\checkmark$ & $\checkmark$ & Identity perception \\
\midrule
\textbf{VIPBench (ours)} & \textbf{Voice} & \textbf{100} & \textbf{9{,}800} & \textbf{124{,}876} & \textbf{1{,}290} & $\checkmark$ & $\checkmark$ & \textbf{Identity perception} \\
\bottomrule
\end{tabular}%
}
\end{table}

\section{The VIPBench dataset}
\label{sec:dataset}

\subsection{Stimuli}

The benchmark contains 9{,}800 voice pairs from 100 contemporary celebrity speakers. The speakers are stratified into 20 cells of five speakers each, formed by crossing five sociophonetic groups (New York City English~\citep{BECKER2014}, Southern American English~\citep{Clopper2005}, African American English~\citep{Wolfram2003}, Latino English~\citep{Wolfram2004}, and Asian American English~\citep{Newman2011}), two genders (male and female), and two age brackets (under 45 and 55 or older). The cells do not cover all sociophonetic variation and given the subjectivity of sociophonetic groupings, these groupings should be treated as coarse sampling strata that enable examining demographic and dialectical breadth, and as such, we use them to balance the speakers over group, gender, and age. The use of public figures' voices for cloning was approved under an IRB protocol (ID: STU00223837). All recordings are conversational speech, with a mean length of 11.6 seconds. The real recordings are excerpts from publicly available interviews and podcasts on YouTube. In each synthetic recording, the cloned or morphed voice speaks a sentence sampled at random from the podcast transcripts in the SPoRC corpus~\citep{litterer-etal-2025-mapping}.

Each pair consists of a reference recording and a comparison recording. Every speaker has one fixed reference recording, which is paired with 98 comparison recordings of six types (Table~\ref{tab:stimtypes}). In different-speaker pairs, the comparison speaker is always one of the other four speakers in the reference speaker's cell. For pairs whose comparison recording is a voice clone, we generated the clone with Cartesia Sonic~3~\citep{cartesia_sonic}, a commercial text-to-speech (TTS) system, using the real comparison recording from the matching real pair as the cloning source.

We created the morphed voices with the voice-mixing feature of Cartesia Sonic~3~\citep{cartesia_sonic}. Each morph mixes the cloned voice of the reference speaker with that of one of the four other speakers in the same cell, and the scale $s$ is the weight given to the reference speaker's voice. Thus, a morph with $s = 0$ is the other speaker's voice alone, and a morph with $s = 1$ is the reference speaker's voice alone. For each reference speaker and each of the four other speakers, we sampled 10 values of $s$ uniformly at random from $[0, 0.98]$, rounded to two decimal places, using a fixed random seed. Each speaker has two source recordings for morphing, and we used the same 10 values of $s$ for both. Together with one unmorphed anchor at $s = 1$, each reference speaker has $4 \times 2 \times 10 + 1 = 81$ morphs, for a total of 8{,}100. Appendix~\ref{sec:generation-app} provides the generation settings and the full sampling procedure, and the release includes the exact value of $s$ for every morph.

The release also includes a held-out extension subset, in which 152 new listeners judged speaker identity for 60 pairs from 10 new speakers, with clones generated by a second TTS system, CosyVoice~2~\citep{du2024cosyvoice2scalablestreaming} (Appendix~\ref{sec:extension-app}).

\begin{table}[t]
\centering
\small
\caption{Stimulus types in VIPBench (1{,}290 listeners, 124{,}876 judgments, 9{,}800 pairs). The comparison recording is paired with the reference recording of the reference speaker. The metadata label is ``same'' for the three same-speaker types and ``different'' for the two different-speaker types, and the morphed voices have no metadata label. Pairs of real recordings have about 39 judgments each, and the pairs with voice clones or morphs have about 10. Reliability is the split-half reliability within each stimulus type (Section~\ref{sec:noise-ceiling}).}
\label{tab:stimtypes}
\setlength{\tabcolsep}{4pt}
\resizebox{\textwidth}{!}{%
\begin{tabular}{cllrrcc}
\toprule
Type & Comparison recording & What the type measures & Pairs & Judgments & Mean $P(\mathrm{same})$ & Reliability \\
\midrule
1 & Same recording, different segment & Ceiling for ``same'' responses & 100 & 3{,}676 & 0.91 & 0.59 \\
2 & Same speaker, different recording & Variation between recordings of a speaker & 400 & 15{,}988 & 0.69 & 0.87 \\
3 & Same speaker, AI voice clone & Whether a clone is heard as its source speaker & 400 & 4{,}024 & 0.59 & 0.58 \\
4 & Different speaker, real recording & Confusion between speakers of the same cell & 400 & 15{,}675 & 0.26 & 0.85 \\
5 & Different speaker, AI voice clone & Confusion with the clone of another speaker & 400 & 4{,}027 & 0.27 & 0.52 \\
6 & Morph of two speakers & Graded targets between two speakers & 8{,}100 & 81{,}486 & 0.38 & 0.66 \\
\bottomrule
\end{tabular}%
}
\end{table}

\subsection{Human annotation protocol}

We recruited 1{,}290 English-speaking listeners through the Centaur AI platform under the same IRB protocol. The platform administers recruitment, panel screening, and performance-based compensation. Listeners did the task on the platform's English web interface, and we set no requirement on listening equipment. Before the task, 1{,}258 listeners answered a questionnaire (Table~\ref{tab:questionnaire} in the appendix). Two thirds of them are female (67\%), and most of them have English as their first language (71\%), are 18 to 34 years old (68\%), and rate their hearing as good or better (90\%).

Each listener completed four practice trials before the main trials. In each trial, the listener heard the reference recording, one second of silence, a half-second beep, one more second of silence, and the comparison recording, and then answered whether the two recordings came from the same speaker. In a second question, the listener was asked ``Who is the first speaker?'' and could choose from four names of speakers from the same speaker cell or ``I don't know''. We call a trial familiar when the chosen name is the reference speaker, and unfamiliar otherwise. Of the 124{,}876 judgments, 78.4\% are from unfamiliar trials (77.7\% to 81.6\% across stimulus types).

Every pair has at least 8 judgments, with a median of 10 and a maximum of 92. The median listener completed 14 trials, and the 10 most active listeners gave 30\% of all judgments. We excluded no listener and no judgment. The 900 pairs of real recordings are gold-standard trials that indicate each listener's performance, and every released judgment on them is marked as correct or incorrect. The only condition on listeners in this paper is a minimum of 10 trials in the individual human baseline (Section~\ref{sec:human-baseline}), which 806 listeners meet.

\subsection{Target definition}

For each pair, we define $P(\mathrm{same}) = n_\mathrm{same} / n_\mathrm{judgments}$, the proportion of listeners who judged the two recordings to come from the same speaker. The target measures perceived sameness. It is a collective judgment of the listeners, and it is not a ground truth about who produced the recordings. In Figure~\ref{fig:psame-by-type} (appendix), most same-recording pairs are above 0.8, and most different-speaker pairs are below 0.5. The perceived sameness of same-speaker clone pairs and morph pairs spans the whole range from 0 to 1.

Listener responses are often different from the metadata label. The metadata label is ``same'' for same-speaker pairs, whose comparison recording is a real recording or a voice clone of the reference speaker. It is ``different'' for different-speaker pairs, whose comparison recording is a real recording or a voice clone of another speaker. On same-speaker pairs, 29\% of the responses are ``different'', and on different-speaker pairs, 26\% of the responses are ``same''. The divergence is largest for same-speaker clones, where 41\% of the responses are ``different''. Counted by pairs, the majority of the listeners disagrees with the metadata label on 19\% of the 900 same-speaker pairs and on 11\% of the 800 different-speaker pairs, and the vote is tied on another 70 and 36 pairs. For same-speaker clones, the majority answers ``different'' on 26\% of the 400 pairs, and the vote is tied on another 16\% (Table~\ref{tab:type-stats} in the appendix).

\subsection{Noise ceiling}
\label{sec:noise-ceiling}

We measure the reliability of $P(\mathrm{same})$ by splitting the listeners into two random halves and correlating the $P(\mathrm{same})$ values of the two halves across pairs. After the Spearman-Brown correction to the full number of listeners, the reliability is $\rho_\mathrm{SB} = 0.705$ (mean over 100 random splits). The reliability is the highest $R^2$ that a model can reach, because the rest of the variance is chance variation between listeners. We call it the noise ceiling (Appendix~\ref{sec:noise-ceiling-app}).

Reliability differs between stimulus types (Table~\ref{tab:stimtypes}). It is highest for the pairs of two different real recordings, which have about 40 judgments each. It is lower for the pairs with a voice clone or a morph, which have about 10 judgments each. The pairs of two segments of one recording have about 37 judgments each, and their $P(\mathrm{same})$ values span a narrow range (SD 0.07, compared with 0.17 to 0.24 for the other types), so their reliability is also low.

\section{Benchmark protocol}
\label{sec:protocol}

We propose four evaluation tasks on the 9{,}800 pairs. In every task, a model's score for a pair is the cosine similarity $c = \mathbf{e}_\mathrm{ref} \cdot \mathbf{e}_\mathrm{comp} / (\|\mathbf{e}_\mathrm{ref}\| \, \|\mathbf{e}_\mathrm{comp}\|)$ between the embeddings of the reference and comparison recordings. When a task requires fitting parameters, such as the calibration in Task~2 or the layer choice for SSL models, we fit them to the listener judgments using 10-fold cross-validation (CV) over speakers. We split the 100 reference speakers into 10 folds of five female and five male speakers each, fit the parameters on the pairs from nine folds, and apply them to the pairs from the held-out fold.

\subsection{Task~1: Continuous $P(\mathrm{same})$ prediction}

Task~1 measures how well a model's cosine similarity $c$ predicts $P(\mathrm{same})$ across all pairs. We report Pearson $r$, $R^2 = r^2$, and $R^2 / \rho_\mathrm{SB}$, which expresses $R^2$ as a proportion of the noise ceiling (Section~\ref{sec:noise-ceiling}). We compute 95\% confidence intervals for $r$ by resampling the pairs with replacement 1{,}000 times. To compare two models, we draw 2{,}000 bootstrap resamples of the pairs and compute the difference between the two models' $r$ values on each resample. The two-sided $p$-value is twice the share of resamples in which the order of the two models is reversed. We correct the $p$-values of all 55 comparisons among the eleven models with the Benjamini-Hochberg procedure~\citep{benjamini1995controlling} at a false discovery rate of 0.05.

\subsection{Task~2: Platt-calibrated verification}

A decision is made between ``same'' and ``different'' only after we set a cutoff on $c$ or turn $c$ into a probability. Task~2 thus measures how well $c$ separates the pairs that most listeners judged to be the same speaker from the pairs that most listeners judged to be different speakers, and how well $c$ can be turned into a probability. We set the label of a pair as $y = 1$ if $P(\mathrm{same}) > 0.5$ and $y = 0$ otherwise. We report the area under the ROC curve (AUC) on the 8{,}884 pairs without a tied vote. For the probability, we fit a logistic regression on $c$ (Platt scaling) with the CV over speakers, which gives a calibrated probability $\hat{p}$ for every pair. We report the expected calibration error (ECE) of $\hat{p}$ over 10 bins with equal numbers of pairs, in which the 916 pairs with a tied vote have $y = 0$. We also report a raw ECE, which compares $c$ itself with $P(\mathrm{same})$ in 10 bins of $c$, without the logistic regression (Appendix~\ref{sec:task2-app}).

\subsection{Task~3: Speaker-level representational similarity}
Studies of voice perception describe the perceived similarity between two voices as a distance in a voice space~\citep{baumann2010perceptual}. Task~3 tests whether the speaker pairs that listeners most often judge to be the same person are also the closest pairs in a model's embedding space. For each reference speaker $A$ and each other speaker $B$ in its cell, the listener dissimilarity is 1 minus the mean $P(\mathrm{same})$ over the two pairs in which $A$'s reference recording is compared with a real recording or a voice clone of $B$. The model dissimilarity is 1 minus the mean $c$ over the same two pairs. The 100 speakers yield 400 ordered pairs $(A, B)$. We calculate a directional dissimilarity matrix from them, analogous to the representational dissimilarity matrix used in representational similarity analysis~\citep{kriegeskorte2008representational}. The only difference is that it is asymmetric, as $(A, B)$ and $(B, A)$ are separate entries computed from different recordings. We report the Spearman correlation between the 400 listener dissimilarities and the 400 model dissimilarities, assessed by a Mantel permutation test (5{,}000 permutations; Appendix~\ref{sec:task3-app}). Same-speaker pairs and morphs are excluded from the matrix; the appendix shows the results for morphs in Figures~\ref{fig:type6-human} and~\ref{fig:type6-models}.

\subsection{Task~4: Real-to-synthetic calibration transfer}
Task~4 evaluates whether the relation between $c$ and $P(\mathrm{same})$ from Task~1 is the same for synthetic recordings as for real recordings. To do this, we fit a line $\hat{P}(\mathrm{same}) = \beta_0 + \beta_1 c$ on the 900 pairs of real recordings and apply it to the 8{,}900 pairs with a voice clone or a morph. We report $R^2_\mathrm{transfer}$, which is 1 minus the ratio of the mean squared error of the predictions to the variance of $P(\mathrm{same})$ on the synthetic pairs. $R^2_\mathrm{transfer}$ is lower when the predictions are farther from the listeners' $P(\mathrm{same})$. We compare it with $R^2_\mathrm{oracle}$, the same quantity for a line fitted only on the synthetic pairs, and we report the ratio of the two slopes, $\beta_1^\mathrm{real} / \beta_1^\mathrm{synth}$. For 95\% intervals, we resample the 100 reference speakers with all their pairs 2{,}000 times and recompute every statistic. The one-sided $p$-value for a slope ratio above 1 is the share of resamples with a ratio of 1 or below, with Benjamini-Hochberg correction across models.

\subsection{Layer selection for models with several layers}
\label{sec:layer-selection}

The models wav2vec~2.0, HuBERT, WavLM, XLS-R, and Whisper give an embedding at every layer, which we average over time. For each fold of the CV, we select the layer whose cosine similarity has the highest Pearson $r$ with $P(\mathrm{same})$ on the nine training folds, and we use the selected layer for the pairs of the held-out fold. The cosine similarities from the selected layers are the scores of the five models in all four tasks. Figure~\ref{fig:layer-alignment} in the appendix shows Pearson $r$ for every layer.

\section{Models}
\label{sec:models}

We evaluate eleven public models without fine-tuning (Table~\ref{tab:models}). Six are supervised speaker models, which are trained with metadata speaker labels. Four are SSL models, and the last one is Whisper, a speech recognition model trained with weak supervision. WeSpeaker ResNet34-LM (WeSpeaker for short) is the speaker embedding model in version~3.1 of the \textit{pyannote} speaker diarization pipeline~\citep{bredin2023pyannote}. We did not evaluate the larger checkpoints of the SSL models and Whisper, or speaker verification models built on SSL features~\citep{chen2022wavlm}.
\begin{table}[t]
\centering
\small
\caption{The eleven models. Dim is the size of the embedding. AAM-Softmax is the additive angular margin softmax loss, and GE2E is the generalized end-to-end loss.}
\label{tab:models}
\resizebox{\textwidth}{!}{%
\begin{tabular}{lllp{5.2cm}r}
\toprule
Model & Type & Training objective & Training data & Dim \\
\midrule
x-vector~\citep{Snyder2018} & Supervised & Softmax & VoxCeleb 1 and 2 & 512 \\
ECAPA-TDNN~\citep{Desplanques_2020} & Supervised & AAM-Softmax & VoxCeleb 1 and 2 & 192 \\
RawNet3~\citep{Jung2022} & Supervised & AAM-Softmax & VoxCeleb 1 and 2 & 192 \\
TitaNet~\citep{Koluguri2022} & Supervised & AAM-Softmax & \raggedright VoxCeleb 1 and 2, Fisher, Switchboard, LibriSpeech, NIST SRE & 192 \\
resemblyzer~\citep{wan2018ge2e} & Supervised & GE2E & LibriSpeech, VoxCeleb 1 and 2 & 256 \\
WeSpeaker~\citep{wang2022wespeakerresearchproductionoriented} & Supervised & AAM-Softmax & VoxCeleb 2 & 256 \\
wav2vec 2.0~\citep{baevski2020wav2vec2} & SSL & Contrastive & LibriSpeech (960 hours) & 768 \\
HuBERT~\citep{hsu2021hubert} & SSL & Masked prediction & LibriSpeech (960 hours) & 768 \\
WavLM~\citep{chen2022wavlm} & SSL & Masked prediction, denoising & \raggedright Libri-Light, GigaSpeech, VoxPopuli (94k hours) & 768 \\
XLS-R~\citep{babu2022xlsr} & SSL & Contrastive & 436k hours in 128 languages & 1024 \\
Whisper~\citep{radford2023whisper} & Weakly supervised & Multitask speech recognition & 680k hours of web audio & 512 \\
\bottomrule
\end{tabular}%
}
\end{table}

Table~\ref{tab:checkpoints} in the appendix lists the model checkpoints and their sources, and every model has 16~kHz mono audio recordings as inputs. Whisper pads every recording to 30~s. We also evaluate the eleven models against the metadata label of each pair (Appendix~\ref{sec:metadata-results}). The six supervised models are trained on VoxCeleb data~\citep{nagrani2017voxceleb,chung2018voxceleb2}, and 26 of our 100 speakers also appear in VoxCeleb by name. On the 5{,}580 pairs with both speakers outside VoxCeleb, the mean $r$ of the supervised models is 0.17 above the mean $r$ of the SSL models, compared with 0.15 on all pairs.
\section{Main results}
\label{sec:main-results}

\subsection{Task~1: Continuous $P(\mathrm{same})$ prediction}

Table~\ref{tab:grand-summary} reports the results for Tasks~1 to~3. Resemblyzer achieves the highest correlation ($r = 0.656$), corresponding to 61\% of the noise ceiling in $R^2$. The other five supervised models range from $r = 0.597$ to $0.646$, the four SSL models range from $0.458$ to $0.510$, and Whisper reaches $0.226$. When we resample the 100 reference speakers, the ranking of model types is unchanged. In all 1{,}000 resamples, every supervised model remains above every SSL model, and every SSL model remains above Whisper. However, the ranking within each model type varies across resamples. For example, resemblyzer, ECAPA-TDNN, and TitaNet do not differ significantly from one another in the speaker-resampling test (Table~\ref{tab:speaker-resampling} in the appendix).

\begin{table}[t]
\centering
\small
\caption{Tasks~1 to~3 for the eleven models, sorted by $r$ within each type. The SSL models and Whisper use the layer selected by CV. The ECE is given without the logistic regression (raw) and with it (cal). $^\dagger$ marks the models in the two comparisons of $r$ that are not significant. The best value of each column is in bold.}
\label{tab:grand-summary}
\begin{tabular}{lccccccc}
\toprule
 & \multicolumn{3}{c}{Task 1} & \multicolumn{3}{c}{Task 2} & Task 3 \\
\cmidrule(lr){2-4} \cmidrule(lr){5-7} \cmidrule(lr){8-8}
Model & $r$ & $R^2$ & $R^2/\rho_\mathrm{SB}$ & AUC & ECE (raw) & ECE (cal) & $\rho$ \\
\midrule
\multicolumn{8}{c}{\emph{Supervised}} \\
resemblyzer & \textbf{0.656} & \textbf{0.430} & \textbf{0.609} & \textbf{0.872} & 0.338 & 0.027 & \textbf{0.542} \\
ECAPA-TDNN & 0.646 & 0.417 & 0.591 & 0.862 & \textbf{0.047} & 0.018 & 0.339 \\
TitaNet & 0.638 & 0.407 & 0.577 & 0.861 & 0.057 & 0.019 & 0.379 \\
WeSpeaker & 0.627 & 0.393 & 0.558 & 0.851 & 0.054 & \textbf{0.012} & 0.389 \\
RawNet3 & 0.617 & 0.381 & 0.540 & 0.851 & 0.086 & 0.021 & 0.386 \\
x-vector & 0.597 & 0.357 & 0.506 & 0.857 & 0.548 & 0.136 & 0.479 \\
\midrule
\multicolumn{8}{c}{\emph{SSL}} \\
WavLM & 0.510 & 0.260 & 0.368 & 0.790 & 0.272 & 0.022 & 0.392 \\
wav2vec 2.0 & 0.471$^\dagger$ & 0.222 & 0.315 & 0.770 & 0.336 & 0.033 & 0.357 \\
XLS-R & 0.465$^\dagger$ & 0.216 & 0.307 & 0.782 & 0.502 & 0.061 & 0.377 \\
HuBERT & 0.458$^\dagger$ & 0.210 & 0.297 & 0.765 & 0.317 & 0.035 & 0.383 \\
\midrule
\multicolumn{8}{c}{\emph{Weakly supervised}} \\
Whisper & 0.226 & 0.051 & 0.072 & 0.651 & 0.594 & 0.037 & 0.127 \\
\bottomrule
\end{tabular}
\end{table}

\subsection{Task~2: Platt-calibrated verification}

The supervised models have an AUC from 0.851 to 0.872, and the SSL models from 0.765 to 0.790. Whisper has 0.651 (Figure~\ref{fig:roc-curves} in the appendix). The raw ECE is above 0.27 for seven models, whose cosine similarities lie far above $P(\mathrm{same})$ on average. After the logistic regression, the ECE is at most 0.061 for every model except x-vector (0.136, Figure~\ref{fig:calibration} in the appendix).

\subsection{Task~3: Speaker-level representational similarity}

The correlation $\rho$ is above zero for every model ($p < 0.05$). Resemblyzer has the highest $\rho$ (0.542) and x-vector the second highest (0.479). The other four supervised models and the four SSL models lie between 0.339 and 0.392, so the supervised models are not above the SSL models as in Task~1. ECAPA-TDNN is second in Task~1 and lowest among these ten models in Task~3. With the entries $(A, B)$ and $(B, A)$ averaged, the order of the models is nearly the same (rank correlation 0.92, Table~\ref{tab:symmetrized} in the appendix).

\subsection{Task~4: Real-to-synthetic calibration transfer}
\label{sec:task4-results}

Table~\ref{tab:task4-transfer} reports Task~4. For every model, $R^2_\mathrm{transfer}$ is below $R^2_\mathrm{oracle}$, and the 95\% interval of the difference is above zero (Table~\ref{tab:task4-gap} in the appendix). The difference is 0.07 to 0.17 for the supervised models. The intervals of $R^2_\mathrm{transfer}$ include zero for wav2vec~2.0, HuBERT, and XLS-R, and the interval of Whisper is below zero. The slope ratio is above 1 for every model ($q \leq 0.014$). For synthetic pairs with a high cosine similarity, listeners therefore answer ``same'' less often than the line fitted on real pairs predicts.

\begin{table}[t]
\centering
\small
\caption{Task~4 with 95\% intervals from 2{,}000 resamples of the reference speakers. The line is fitted on the 900 pairs of real recordings and applied to the 8{,}900 pairs with a voice clone or a morph. $q$ is the corrected $p$-value for a slope ratio above 1. Models are sorted by $R^2_\mathrm{transfer}$ within each type.}
\label{tab:task4-transfer}
\begin{tabular}{lcccc}
\toprule
Model & $R^2_\mathrm{transfer}$ & $R^2_\mathrm{oracle}$ & Slope ratio & $q$ \\
\midrule
\multicolumn{5}{c}{\emph{Supervised}} \\
resemblyzer & 0.297 [0.242, 0.345] & 0.379 [0.346, 0.411] & 1.29 [1.21, 1.38] & $<$0.001 \\
ECAPA-TDNN & 0.282 [0.223, 0.332] & 0.362 [0.328, 0.395] & 1.13 [1.06, 1.19] & $<$0.001 \\
WeSpeaker & 0.268 [0.211, 0.317] & 0.335 [0.300, 0.371] & 1.15 [1.07, 1.22] & $<$0.001 \\
TitaNet & 0.266 [0.208, 0.317] & 0.351 [0.316, 0.385] & 1.16 [1.09, 1.24] & $<$0.001 \\
RawNet3 & 0.184 [0.114, 0.243] & 0.330 [0.293, 0.365] & 1.22 [1.15, 1.30] & $<$0.001 \\
x-vector & 0.141 [0.064, 0.206] & 0.312 [0.281, 0.343] & 1.64 [1.50, 1.79] & $<$0.001 \\
\midrule
\multicolumn{5}{c}{\emph{SSL}} \\
WavLM & 0.131 [0.062, 0.190] & 0.223 [0.189, 0.258] & 1.63 [1.45, 1.83] & $<$0.001 \\
wav2vec 2.0 & 0.045 [$-0.026$, 0.110] & 0.192 [0.158, 0.229] & 1.66 [1.50, 1.85] & $<$0.001 \\
HuBERT & 0.039 [$-0.034$, 0.102] & 0.179 [0.145, 0.217] & 1.70 [1.51, 1.93] & $<$0.001 \\
XLS-R & 0.003 [$-0.079$, 0.074] & 0.186 [0.155, 0.218] & 1.88 [1.68, 2.10] & $<$0.001 \\
\midrule
\multicolumn{5}{c}{\emph{Weakly supervised}} \\
Whisper & $-0.146$ [$-0.272$, $-0.093$] & 0.046 [0.026, 0.069] & 2.10 [1.16, 3.40] & 0.014 \\
\bottomrule
\end{tabular}
\end{table}

\subsection{Learned similarity functions, layers, and robustness}
\label{sec:further-analyses}

\textbf{Learned similarity functions.} We also fitted three learned similarity functions with the CV over speakers (Tables~\ref{tab:mahalanobis} and~\ref{tab:probes} in the appendix). A diagonal Mahalanobis metric, which learns one weight for each dimension of the embedding, changes $R^2$ by $-0.03$ to $+0.01$ for the supervised models, by $+0.02$ to $+0.03$ for the SSL models, and by $+0.10$ for Whisper. A ridge regression on pair features reduces the difference in $r$ between the weakest supervised model and the strongest SSL model from 0.09 to 0.02. The order of the three types of models is the same for every function.

\textbf{Layers of the SSL models.} For the four SSL models, $r$ is highest at one of the first two layers and lower at later layers (Figure~\ref{fig:layer-alignment} in the appendix). With the last layer, the four models have an $r$ from 0.03 to 0.27, compared with 0.46 to 0.51 with the selected layer. The selected layer is the same in every fold for every model except HuBERT, which alternates between layers~0 and~1 (Table~\ref{tab:per-fold-layer}).

\textbf{Robustness.} The ranking in Task~1 is identical when we compute $P(\mathrm{same})$ from unfamiliar trials alone, and the best model then reaches 63\% of the noise ceiling of that target (Table~\ref{tab:unfamiliar} in the appendix). The ranking is also identical without the 10 most active listeners. The appendix reports the ranking on real recordings alone and with equal weight for each stimulus type (Table~\ref{tab:robustness}), on the pairs outside VoxCeleb (Table~\ref{tab:voxceleb}), and on the CosyVoice~2 extension subset (Table~\ref{tab:extension-results}).

\section{Individual human baseline}
\label{sec:human-baseline}

For each of the 806 listeners with at least 10 trials, we compute how often the listener agrees with the majority of the other listeners on the pairs that the listener judged. The mean agreement is 0.73 (median 0.74, SD 0.12). For each model, we set a cutoff on the cosine similarity with the CV over speakers, and we compute how often the model agrees with the majority of all listeners on the 8{,}884 pairs without a tied vote. Resemblyzer agrees on 0.79 of the pairs, which is more than 72\% of the listeners reach. The other five supervised models are above 65\% to 70\% of the listeners, and the SSL models above 33\% to 39\% of the listeners. Whisper is above 19\% of the listeners (Table~\ref{tab:nested-thresholds} in the appendix).

\section{Demographic fairness}
\label{sec:fairness}

Speech recognition systems have higher error rates for African American speakers than for white speakers~\citep{koenecke2020racial}. We therefore compare $r$ in Task~1 between the pairs of female and of male speakers, and between the pairs of the five sociophonetic groups (Figure~\ref{fig:fairness} in the appendix). For every model except Whisper, the $r$ for female speakers and the $r$ for male speakers differ by at most 0.014. For Whisper the difference is 0.12, and the lower $r$ is for female speakers. Across the five sociophonetic groups, the difference between the highest and the lowest $r$ is 0.11 to 0.17 for every model, and the same group has the lowest $r$ for ten of the eleven models. A group has 20 speakers, so we do not test the differences between groups.

\section{Discussion}

\subsection{Recommendations}

\textbf{Recommendations for VIPBench evaluations.} We recommend that evaluations (1) report all four tasks, because a model's rank can differ across tasks (Section~\ref{sec:main-results}); (2) calibrate the cosine similarity as in Task~2 whenever a confidence estimate is needed; (3) for SSL models, use CV to select the layer that best represents speaker identity instead of defaulting to the last layer; and (4) when human perceptual judgments are the target, report results for real and synthetic recordings separately, because a regression fitted to listener judgments on real pairs does not fully transfer to synthetic pairs.

\textbf{Research directions.} Our results suggest three research directions. First, none of the models we evaluate reaches the noise ceiling of human perception, and none was trained on listener judgments. Speaker models trained on human perceptual judgments may close the gap towards the ceiling, which would make them useful in human--AI collaboration tasks. Second, calibration fitted on real pairs does not fully transfer to synthetic pairs, so calibration and training methods that generalize from real to synthetic voices would be useful in settings where voice clones are common. Third, the weak correlation between model confidence and listener disagreement (at most $r = 0.34$) leaves human-aligned uncertainty estimation as an open problem.

\subsection{Limitations}
VIPBench includes 100 speakers, all public figures who speak American English, and its real recordings come from interviews and podcasts. The benchmark therefore does not cover other languages or recording conditions, and because it includes no speakers who are not public figures, we cannot compare how listeners perceive the voices of public figures and private individuals. All voice clones and morphs in the main set come from a single commercial system, so the magnitude of the slope ratio in Task~4 may be specific to that system, and the CosyVoice~2 subset is too small to resolve this question (Table~\ref{tab:extension-results}). No listener judged the same pair twice, so we cannot estimate test-retest reliability for individual listeners (Appendix~\ref{sec:consistency-app}).

\subsection{Ethics, access, and responsible use}

Knowledge of which voice clones listeners accept could help someone make fraudulent calls more convincing. The same knowledge helps defenses, because a model that predicts the judgments of listeners can find the clones that listeners accept. We release the benchmark on Hugging Face\footnote{\url{https://huggingface.co/datasets/sendfuze/vipbench}} with Croissant metadata, which includes the Responsible AI fields. We do not release guidance for making voice clones that listeners accept more often. The datasheet documents the generation settings that are needed to verify the stimuli. We generated all voice clones from public recordings of public figures under an IRB protocol (Section~\ref{sec:dataset}), and the listeners gave consent on the Centaur platform.

\section*{Acknowledgments and Disclosure of Funding}
We gratefully acknowledge comments and feedback from Erik Duhaime and Michelle Morgenthal at Centaur AI, and funding and support from VeryAI. The authors declare no competing interests.

\bibliographystyle{unsrtnat}
\bibliography{references}

\FloatBarrier
\newpage
\appendix

\section{Technical appendices and supplementary material}

\paragraph{Dataset and code release.} We release the dataset on Hugging Face at \url{https://huggingface.co/datasets/sendfuze/vipbench}. The release has the audio files, the individual judgments with the questionnaire answers of the listeners, the embeddings of the eleven models, and the embeddings at every layer for the four SSL models and Whisper. It also has a datasheet and Croissant metadata with the Responsible AI fields. A sample with five speakers and all six stimulus types (about 290\,MB) is at \url{https://huggingface.co/datasets/sendfuze/vipbench/tree/main/samples}. The extraction scripts and the analysis notebook are in the folder \texttt{code/} of the same repository, and the notebook runs on the released embeddings without the audio.

\paragraph{Repository map.}
\label{sec:repository-map}
Table~\ref{tab:repository-map} lists the file behind each input of the analyses and the analyses that use it. The analysis notebook builds the cross-validation folds from the speaker list with a fixed seed. It shuffles the 50 male and the 50 female speaker IDs with seed 42 and deals them into 10 folds of five male and five female speakers.

\begin{table}[h]
\centering
\small
\caption{Files of the release and the analyses that use them. Paths are relative to the root of the dataset repository.}
\label{tab:repository-map}
\begin{tabular}{p{4.1cm}p{5.2cm}p{3.3cm}}
\toprule
Input & File & Used for \\
\midrule
Speakers with sociophonetic group, gender, and age bracket & \path|data/speakers.csv| & Fairness analysis, CV folds, Task~3 \\
Pairs with stimulus type, metadata label, and vote counts & \path|data/stimuli.csv| & All four tasks, analyses by stimulus type \\
Exact morph weights and transcripts & \path|data/stimuli_interpol.csv| & Analyses along the morph weight \\
Individual judgments, familiarity flag, and questionnaire answers & \path|data/participant_responses.csv| & $P(\mathrm{same})$, noise ceiling, unfamiliar trials, human baseline \\
Embeddings of the eleven models & \path|data/embeddings/{model}.npz| & All four tasks \\
Embeddings at every layer for the SSL models and Whisper & \path|data/embeddings/layers/{model}.npz| & Layer selection, layer analysis \\
Audio & \path|data/audio/reference/|, \path|data/audio/comparison/| & Extracting embeddings \\
Analysis notebook & \path|code/benchmark_analysis.ipynb|, run by \path|code/reproduce.sh| & Tables and figures of the four tasks \\
Extraction scripts & \path|code/extract_*.py|, \path|code/run_all_extractions.sh| & Rebuilding the embeddings \\
CosyVoice~2 extension subset & \path|data/experiment2/|, \path|code/extension_cosyvoice2.py| & Results on the extension subset \\
Sample of five speakers & \path|samples/| & Quick inspection \\
\bottomrule
\end{tabular}
\end{table}

\paragraph{Compute.} We extracted the embeddings on one laptop with an NVIDIA RTX 4070 GPU (8\,GB of memory). The extraction of all models took about 24 CPU-hours and about one GPU-hour for Whisper. Users do not need to repeat the extraction, because the release includes the embeddings. The analysis notebook runs in about 10 minutes on the released embeddings.

\paragraph{Stimulus generation.}
\label{sec:generation-app}
We generated all clones and morphs with the Cartesia API. For each clone, we created a cloned voice from one real source recording with the voice-cloning endpoint in its ``similarity'' mode, with enhancement turned off, and we synthesized the speech with the model \texttt{sonic-3} as 16\,kHz audio. The source recording of a clone is the real comparison recording of the matching real pair. The datasheet in the release lists every setting.

For a morph, we mixed two cloned voices with the voice-mixing endpoint, with weight $s$ on the reference speaker's voice and $1 - s$ on the comparison speaker's voice. Each speaker has two source recordings for morphing, A and B, and a morph mixes the two speakers' clones of recording~A or their clones of recording~B. For each reference speaker and each of its four comparison speakers, we drew 10 values of $s$ uniformly from 0 to 0.98 at two decimals (Python \texttt{random}, seed 42), and both recordings used the same 10 values. Some draws repeated a value, and some pairs of values produced the same file name. We gave each affected morph a new value drawn from the unused two-decimal values between 0 and 0.99. After the replacement, 354 of the 8{,}000 morphs without the anchors have a value that the other recording of the same speaker pair does not use, and 4 morphs have $s = 0.99$. The anchor of a reference speaker is the speaker's clone of recording~A mixed with itself at $s = 1$. The file \texttt{stimuli\_interpol.csv} lists the exact $s$ and the transcript of every morph. The \texttt{scale} column of \texttt{stimuli.csv} holds the integer in the file name, which is one point below $100s$ for 245 morphs because of floating-point truncation.

\begin{figure}[h]
\centering
\includegraphics[width=0.55\linewidth]{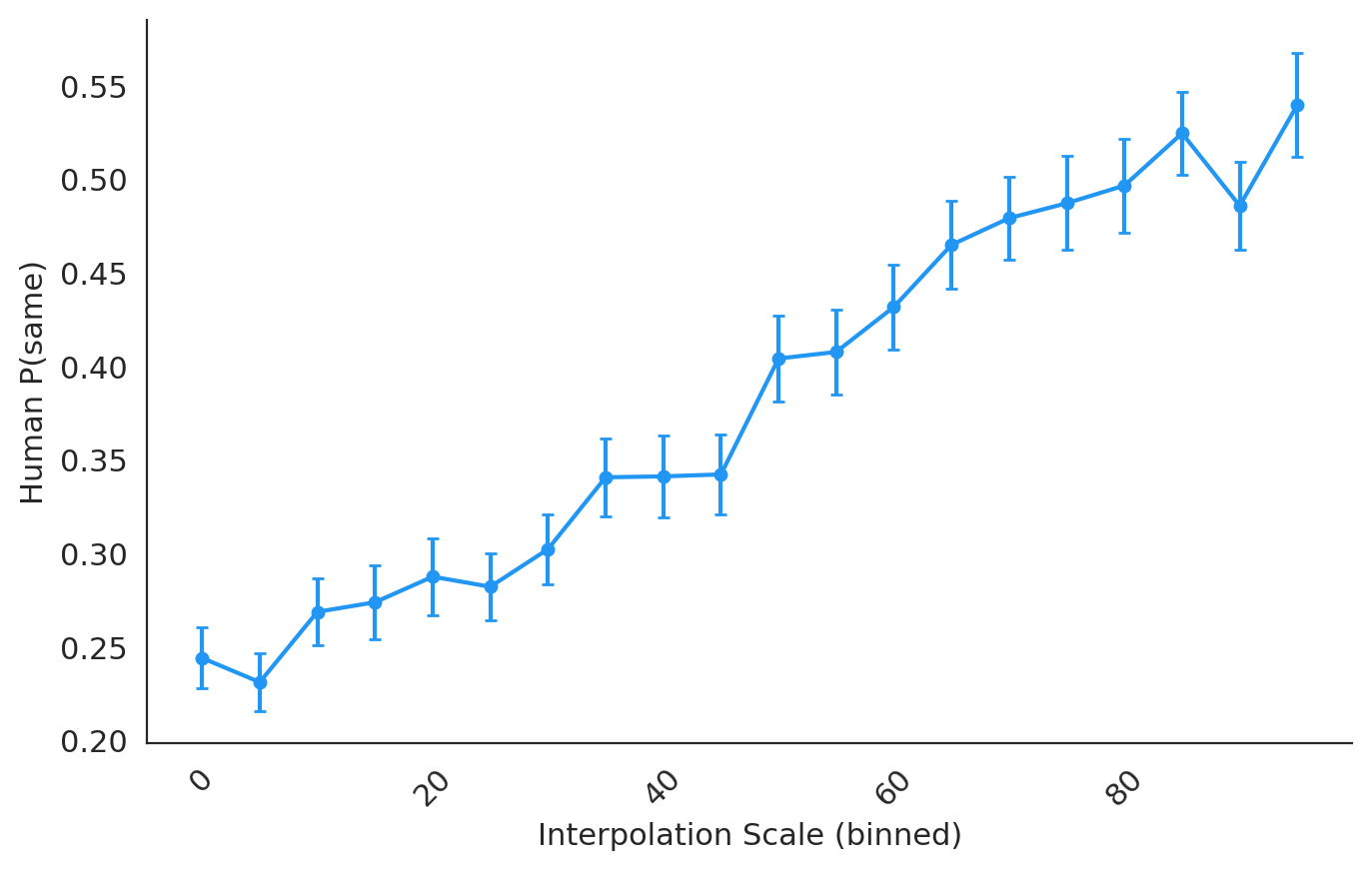}
\caption{$P(\mathrm{same})$ as a function of the morph weight $s$ (in percent, binned). Error bars are 1.96$\times$ SEM in each bin.}
\label{fig:type6-human}
\end{figure}

Mean $P(\mathrm{same})$ rises with $s$ (Figure~\ref{fig:type6-human}). It is 0.24 for $s$ up to 0.10, 0.37 for $s$ from 0.46 to 0.55, 0.50 for $s$ from 0.91 to 0.99, and 0.55 for the anchors. Of the 9{,}900 recordings, 87 are shorter than 2 seconds (85 morphs with short sampled sentences and 2 real recordings of different speakers), and the shortest recording is 0.09 seconds.

\paragraph{CosyVoice~2 extension subset.}
\label{sec:extension-app}
The extension subset is a held-out set with clones from a second TTS system. It has 10 public figures who are not among the 100 speakers of the main set, and each of them is paired with one soundalike, a different public figure chosen for a similar voice. For each of the 10 speakers, listeners compared the reference recording with a different recording of the speaker, with two clones of the speaker, with a recording of the soundalike, and with two clones of the soundalike. The six pairs correspond to the four types of the main set with a real recording or a voice clone of the same speaker or of a different speaker. We generated the clones with CosyVoice~2~\citep{du2024cosyvoice2scalablestreaming} in its zero-shot mode, and the voice prompt of each clone is the real comparison recording of the matched pair. One of the two clones speaks the words of the reference recording, and the other speaks a sentence sampled from the SPoRC corpus. We recruited 152 new listeners on Prolific under the same IRB determination, and they gave 1{,}291 judgments on the 60 pairs, with a median of 21 judgments for each pair. Listeners answered on a four-point scale from ``definitely different'' to ``definitely the same'', and we count the two upper points as ``same'' when we compute $P(\mathrm{same})$.

We applied the eleven models to the 60 pairs without any fitting, and the SSL models and Whisper use the layer that was selected most often on the main set. Table~\ref{tab:extension-results} reports $r$ with 95\% intervals from 2{,}000 resamples of the 10 speakers. The rank correlation between the ranking on the subset and the ranking on the main set is 0.73, with a 95\% interval from $-0.16$ to 0.82. The mean $r$ of the supervised models (0.53) is above the mean $r$ of the SSL models (0.28), and Whisper has the lowest $r$ (0.15). Every supervised model is above every SSL model and every SSL model is above Whisper in 49\% of the resamples. The slope ratio of Task~4, computed with the real pairs and the clone pairs of the subset, is above 1 for the six supervised models (1.40 to 1.98), and its interval includes 1 for every model. The subset has 60 pairs from 10 speakers, so it can show only large differences between models.

\begin{table}[h]
\centering
\small
\caption{$r$ on the main set and on the CosyVoice~2 extension subset, with 95\% intervals from resamples of the 10 speakers of the subset.}
\label{tab:extension-results}
\begin{tabular}{lcc}
\toprule
Model & Main set & Extension subset \\
\midrule
\multicolumn{3}{c}{\emph{Supervised}} \\
resemblyzer & 0.656 & 0.51 [0.37, 0.66] \\
ECAPA-TDNN & 0.646 & 0.50 [0.06, 0.79] \\
TitaNet & 0.638 & 0.49 [0.02, 0.84] \\
WeSpeaker & 0.627 & 0.49 [$-0.01$, 0.86] \\
RawNet3 & 0.617 & 0.56 [0.17, 0.80] \\
x-vector & 0.597 & 0.65 [0.52, 0.77] \\
\midrule
\multicolumn{3}{c}{\emph{SSL}} \\
WavLM & 0.510 & 0.30 [0.08, 0.54] \\
wav2vec 2.0 & 0.471 & 0.23 [0.07, 0.50] \\
XLS-R & 0.465 & 0.33 [0.14, 0.53] \\
HuBERT & 0.458 & 0.24 [0.03, 0.51] \\
\midrule
\multicolumn{3}{c}{\emph{Weakly supervised}} \\
Whisper & 0.226 & 0.15 [0.01, 0.30] \\
\bottomrule
\end{tabular}
\end{table}

\paragraph{Listeners.}
\label{sec:listeners-app}
Table~\ref{tab:questionnaire} lists the questionnaire answers of 1{,}258 listeners. The other 32 listeners did not complete the questionnaire. Each trial was one video with an animated waveform. Its audio track held the reference recording, one second of silence, a half-second 440\,Hz beep, one more second of silence, and the comparison recording. The identity question was ``Are these two clips from the same speaker?'', with the answers ``Same'' and ``Different''.

The number of trials for each listener ranges from 1 to 5{,}509, with a median of 14. The 10 most active listeners gave 30\% of the 124{,}876 judgments, the 20 most active gave 46\%, and the 100 most active gave 77\%. We recomputed $P(\mathrm{same})$ without the judgments of the 10 most active listeners, and again without the judgments of the 20 most active listeners. In both cases, the Task~1 ranking of the eleven models is identical to the ranking on all judgments.

\begin{table}[h]
\centering
\small
\caption{Questionnaire answers of the 1{,}258 listeners who completed the questionnaire. Each entry is a number of listeners.}
\label{tab:questionnaire}
\begin{tabular}{lp{10.8cm}}
\toprule
Question & Answers \\
\midrule
Gender & 837 female, 398 male, 23 ``prefer not to say'' or ``intersex'' \\
Age & 25 under 18, 451 aged 18 to 24, 407 aged 25 to 34, 192 aged 35 to 44, 101 aged 45 to 54, 57 aged 55 to 64, 6 aged 65 or older, 19 ``prefer not to say'' \\
First language & 899 English, 323 another language, 36 ``prefer not to say'' \\
Hearing & 17 poor, 104 fair, 352 good, 379 very good, 406 excellent \\
\bottomrule
\end{tabular}
\end{table}

\paragraph{Unfamiliar trials.}
\label{sec:unfamiliar-app}
We recomputed $P(\mathrm{same})$ from the 97{,}906 judgments of unfamiliar trials and repeated Tasks~1 and~2 (Table~\ref{tab:unfamiliar}). Every pair has at least one such judgment, and the median is 8. The Task~1 ranking and the Task~2 AUC ranking of the eleven models are both identical to the rankings on all judgments. Pearson $r$ is lower by at most 0.033, and AUC is lower by 0.015 to 0.025. The split-half reliability of the unfamiliar target is 0.633, so the best model reaches 63\% of the noise ceiling on unfamiliar trials and 61\% on all judgments. We also computed $P(\mathrm{same})$ from familiar trials alone, for the 4{,}291 pairs with at least three familiar judgments. The Task~1 ranking on those pairs has a rank correlation of 0.94 with the ranking on all judgments. Every supervised model stays above every SSL model, and every SSL model stays above Whisper.

\begin{table}[h]
\centering
\small
\caption{Tasks~1 and~2 with $P(\mathrm{same})$ computed from all judgments, from unfamiliar trials alone, and from familiar trials alone. The familiar column uses the 4{,}291 pairs with at least three familiar judgments. AUC leaves out pairs with a tied vote. WeSpeaker has no score for one pair, whose comparison recording is 0.09 seconds long.}
\label{tab:unfamiliar}
\begin{tabular}{lccccc}
\toprule
 & \multicolumn{3}{c}{Task~1 Pearson $r$} & \multicolumn{2}{c}{Task~2 AUC} \\
\cmidrule(lr){2-4}\cmidrule(lr){5-6}
Model & All & Unfamiliar & Familiar & All & Unfamiliar \\
\midrule
resemblyzer & 0.656 & 0.631 & 0.560 & 0.872 & 0.850 \\
ECAPA-TDNN  & 0.646 & 0.613 & 0.576 & 0.862 & 0.840 \\
TitaNet     & 0.638 & 0.606 & 0.572 & 0.861 & 0.837 \\
WeSpeaker   & 0.627 & 0.594 & 0.567 & 0.851 & 0.828 \\
RawNet3     & 0.617 & 0.585 & 0.555 & 0.851 & 0.827 \\
x-vector    & 0.597 & 0.572 & 0.525 & 0.857 & 0.835 \\
WavLM       & 0.510 & 0.490 & 0.445 & 0.790 & 0.771 \\
wav2vec 2.0 & 0.471 & 0.453 & 0.415 & 0.770 & 0.753 \\
XLS-R       & 0.465 & 0.446 & 0.415 & 0.782 & 0.763 \\
HuBERT      & 0.458 & 0.440 & 0.400 & 0.765 & 0.751 \\
Whisper     & 0.226 & 0.210 & 0.219 & 0.651 & 0.634 \\
\bottomrule
\end{tabular}
\end{table}

\begin{figure}[h]
\centering
\includegraphics[width=\linewidth]{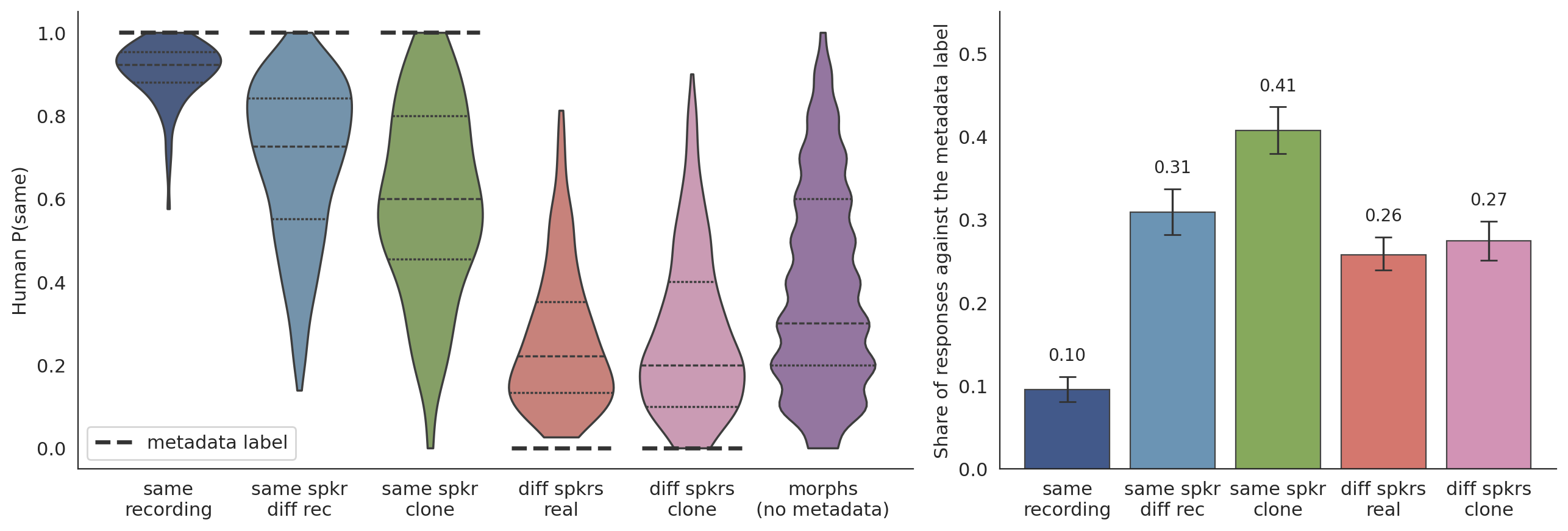}
\caption{\textbf{Left.} Distribution of $P(\mathrm{same})$ for each stimulus type over all 1{,}290 listeners. The thick dashed segment marks the metadata label (``same'' for the three same-speaker types and ``different'' for the two different-speaker types), and the lines inside each violin mark the quartiles. \textbf{Right.} Share of listener responses that contradict the metadata label, with 95\% bootstrap intervals over reference speakers. The share is largest for same-speaker AI voice clones (41\%).}
\label{fig:psame-by-type}
\end{figure}

\paragraph{Statistics by stimulus type.}
\label{sec:type-stats-app}
Figure~\ref{fig:psame-by-type} shows the distribution of $P(\mathrm{same})$ for each stimulus type and the share of listener responses that contradict the metadata label. Table~\ref{tab:type-stats} lists the reliability of each type and three measures of the divergence from the metadata label. We computed the reliability of a type with the split-half procedure of Section~\ref{sec:noise-ceiling}, applied to the pairs of the type. A pair is tied when exactly half of its judgments are ``same''. Tied pairs are more frequent in the types with about 10 judgments for each pair, so we compare types by the share of responses, which needs no rule for tied pairs.

\begin{table}[h]
\centering
\small
\caption{Statistics of each stimulus type. The split-half correlation is the mean over 100 random splits of the listeners, and the reliability is its Spearman-Brown correction. The last three columns compare the listener responses with the metadata label, which the morphs do not have.}
\label{tab:type-stats}
\setlength{\tabcolsep}{4pt}
\begin{tabular}{crccccccc}
\toprule
 & & Judgments & SD of & Split-half & & Responses & Pairs with majority & Tied \\
Type & Pairs & per pair & $P(\mathrm{same})$ & correlation & Reliability & against label & against label & pairs \\
\midrule
1 & 100 & 36.8 & 0.07 & 0.42 & 0.59 & 9.5\% & 0.0\% & 0.0\% \\
2 & 400 & 40.0 & 0.19 & 0.76 & 0.87 & 30.9\% & 17.3\% & 1.8\% \\
3 & 400 & 10.1 & 0.23 & 0.41 & 0.58 & 40.7\% & 26.0\% & 15.8\% \\
4 & 400 & 39.2 & 0.17 & 0.74 & 0.85 & 25.8\% & 11.5\% & 0.5\% \\
5 & 400 & 10.1 & 0.19 & 0.35 & 0.52 & 27.4\% & 10.5\% & 8.5\% \\
6 & 8{,}100 & 10.1 & 0.24 & 0.49 & 0.66 & n/a & n/a & 10.0\% \\
\bottomrule
\end{tabular}
\end{table}

\paragraph{Noise-ceiling derivation.}
\label{sec:noise-ceiling-app}
We split the 1{,}290 listeners into two random halves, compute $P(\mathrm{same})$ for every pair in each half, and take the Pearson correlation $r_\mathrm{half}$ between the two sets of values. The Spearman-Brown formula $\rho_\mathrm{SB} = 2 r_\mathrm{half} / (1 + r_\mathrm{half})$ gives the reliability for the full number of listeners. The reported value of 0.705 is the mean over 100 random splits. A model that predicted the $P(\mathrm{same})$ of an unlimited number of listeners exactly would reach $r = \sqrt{\rho_\mathrm{SB}} = 0.84$ and $R^2 = \rho_\mathrm{SB} = 0.705$ against the observed $P(\mathrm{same})$, so these values are the upper bounds of the two metrics.

\begin{figure}[h]
\centering
\includegraphics[width=0.95\linewidth]{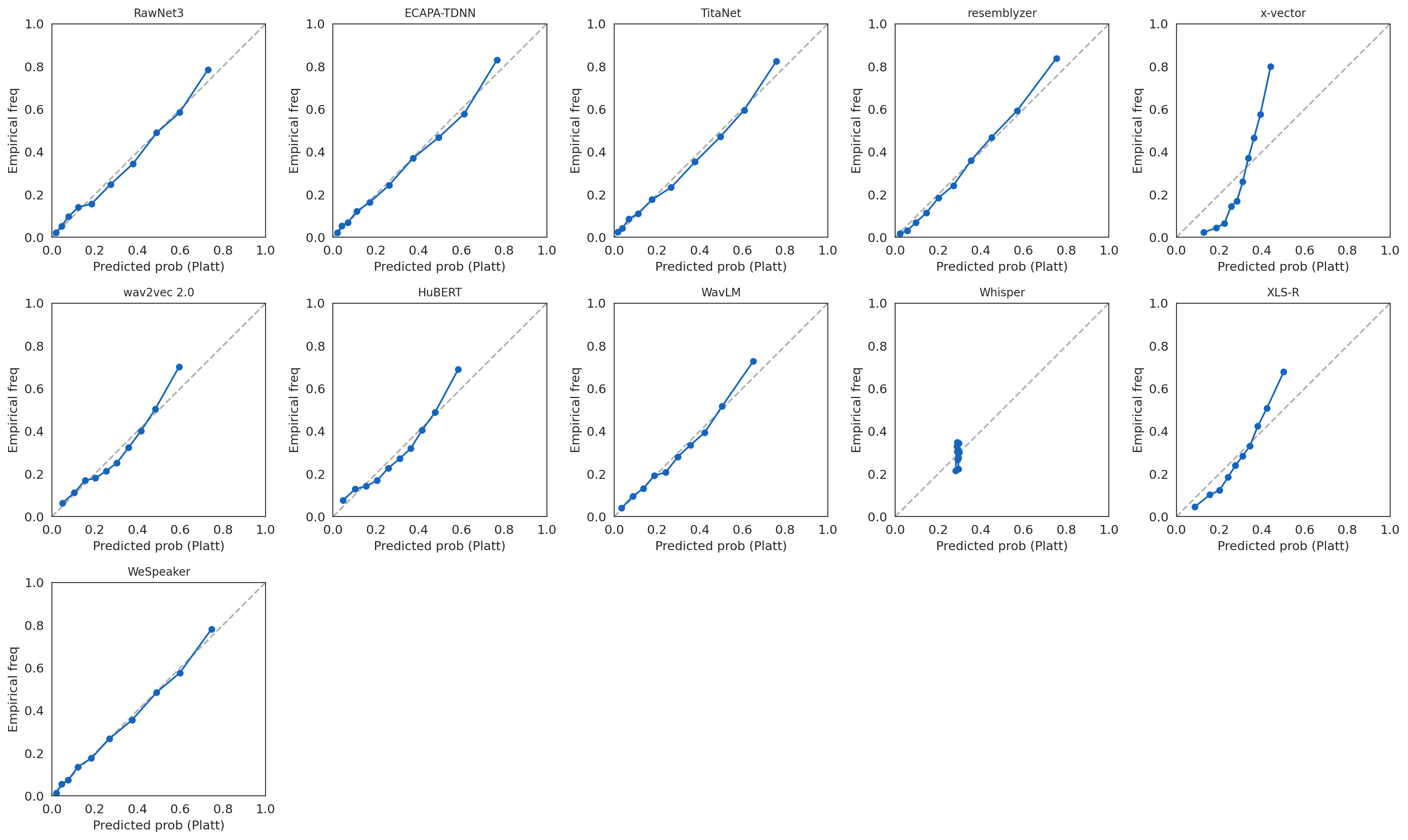}
\caption{Reliability diagrams after Platt scaling. Each panel plots the mean $\hat{p}$ of a bin against the share of pairs in the bin whose majority answer is ``same'', for 10 bins with equal numbers of pairs. A model on the diagonal is perfectly calibrated. All models are close to the diagonal except x-vector, whose calibrated ECE is 0.136.}
\label{fig:calibration}
\end{figure}

\paragraph{Task~2 calibration details.}
\label{sec:task2-app}
Platt scaling fits, in each fold,
\begin{equation*}
\hat{p} = \sigma(\alpha \cdot c + \beta), \qquad \sigma(z) = \frac{1}{1 + e^{-z}},
\end{equation*}
to the label $y$ of the pairs of the nine training folds, and the $\hat{p}$ of a pair comes from the fold in which its reference speaker is held out. The calibrated ECE uses 10 bins of $\hat{p}$ with equal numbers of pairs,
\begin{equation*}
\mathrm{ECE}_\mathrm{cal} = \sum_{b=1}^{10} \frac{n_b}{N} \left| \bar{p}_b - \bar{y}_b \right|,
\end{equation*}
where $n_b$ is the number of pairs in bin $b$, $N = 9{,}800$, $\bar{p}_b$ is the mean $\hat{p}$ in bin $b$, and $\bar{y}_b$ is the mean label in bin $b$. The raw ECE uses 10 bins of $c$, and it replaces $\bar{p}_b$ with the mean $c$ and $\bar{y}_b$ with the mean $P(\mathrm{same})$ of the bin. For seven models, the mean $c$ is above the mean $P(\mathrm{same})$ in every bin, so their raw ECE equals the mean $c$ minus the mean $P(\mathrm{same})$ over all pairs (0.398). Figure~\ref{fig:calibration} shows the reliability diagrams after Platt scaling.

\begin{figure}[h]
\centering
\includegraphics[width=\linewidth]{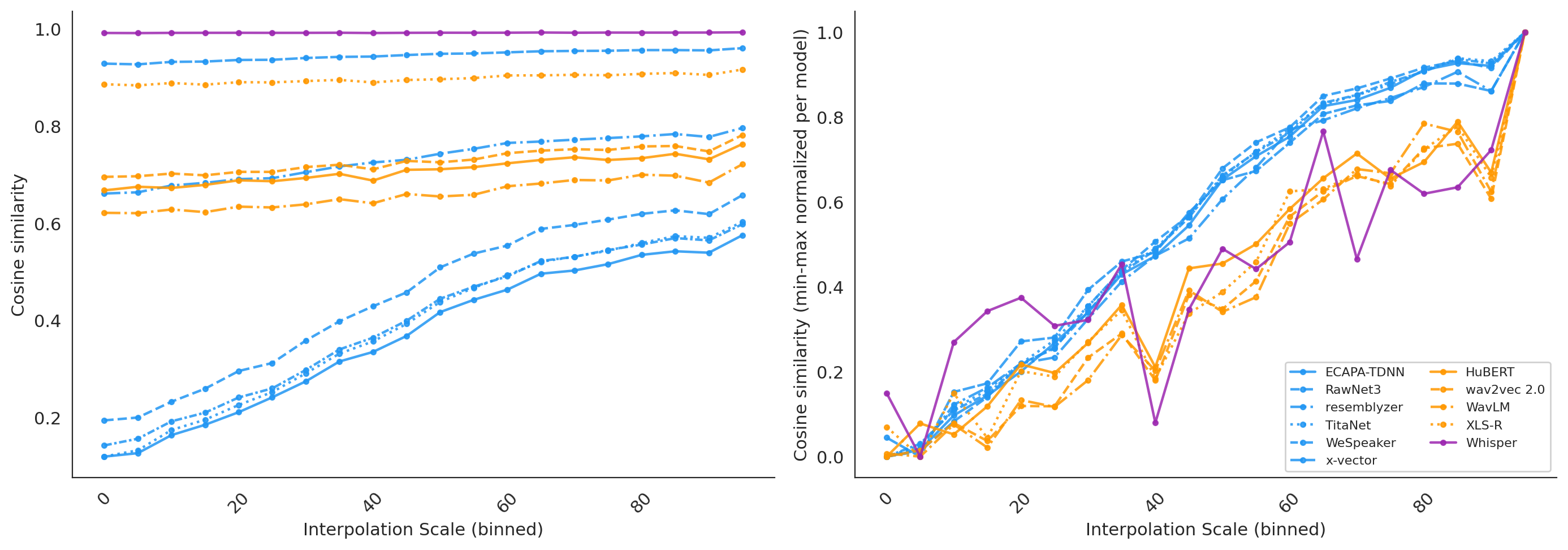}
\caption{Cosine similarity of each model along the morph weight $s$. \textbf{Left.} Raw cosine similarity. \textbf{Right.} Cosine similarity scaled to the range from 0 to 1 within each model. The colors mark the type of the model (blue for supervised, orange for SSL, and purple for weakly supervised).}
\label{fig:type6-models}
\end{figure}

\paragraph{Task~3 details and the symmetrized matrix.}
\label{sec:task3-app}
The listener matrix has the entries $1 - \overline{P(\mathrm{same})}$, and the model matrix has the entries $1 - \overline{c}$, where the mean is over the two pairs of the reference speaker $A$ and the comparison speaker $B$. Each matrix has 400 entries, because every reference speaker is paired only with the four other speakers of its cell. We leave out the same-speaker pairs. The mean listener dissimilarity is 0.09 to 0.41 for the three same-speaker types and 0.73 to 0.74 for the two different-speaker types, so the same-speaker pairs would raise $\rho$ through the difference between the two groups of pairs. The permutation test shuffles the 400 model values 5{,}000 times, and the $p$-value is the share of shuffles with $|\rho|$ at least as large as the observed $|\rho|$. It is below 0.001 for ten models and 0.009 for Whisper. When we permute the speakers instead, so that entries with a shared speaker move together, the $p$-value stays below 0.001 for the ten models and is 0.031 for Whisper. Table~\ref{tab:symmetrized} reports $\rho$ for the symmetrized matrix, which averages the entries $(A, B)$ and $(B, A)$ and has 200 entries. Every $\rho$ rises, by at most 0.07, and the rank correlation between the two orderings of the models is 0.92. For the symmetrized matrix, the two tests give $p = 0.026$ and $p = 0.065$ for Whisper and $p < 0.001$ for the other models. Figures~\ref{fig:type6-human} and~\ref{fig:type6-models} show $P(\mathrm{same})$ and the cosine similarity along the morph weight.

\begin{table}[h]
\centering
\small
\caption{Task~3 $\rho$ for the directional matrix (400 entries) and the symmetrized matrix (200 entries).}
\label{tab:symmetrized}
\begin{tabular}{lcc}
\toprule
Model & Directional & Symmetrized \\
\midrule
\multicolumn{3}{c}{\emph{Supervised}} \\
resemblyzer & 0.542 & 0.615 \\
x-vector & 0.479 & 0.505 \\
WeSpeaker & 0.389 & 0.439 \\
RawNet3 & 0.386 & 0.421 \\
TitaNet & 0.379 & 0.433 \\
ECAPA-TDNN & 0.339 & 0.350 \\
\midrule
\multicolumn{3}{c}{\emph{SSL}} \\
WavLM & 0.392 & 0.436 \\
HuBERT & 0.383 & 0.442 \\
XLS-R & 0.377 & 0.384 \\
wav2vec 2.0 & 0.357 & 0.379 \\
\midrule
\multicolumn{3}{c}{\emph{Weakly supervised}} \\
Whisper & 0.127 & 0.158 \\
\bottomrule
\end{tabular}
\end{table}

\paragraph{Task~4 intervals.}
\label{sec:task4-app}
Table~\ref{tab:task4-gap} reports the difference $R^2_\mathrm{oracle} - R^2_\mathrm{transfer}$ and the difference between the two slopes, $\beta_1^\mathrm{real} - \beta_1^\mathrm{synth}$, with 95\% intervals from the 2{,}000 resamples of the reference speakers. Both intervals are above zero for every model.

\begin{table}[h]
\centering
\small
\caption{Task~4 differences with 95\% intervals from 2{,}000 resamples of the reference speakers.}
\label{tab:task4-gap}
\begin{tabular}{lcc}
\toprule
Model & $R^2_\mathrm{oracle} - R^2_\mathrm{transfer}$ & $\beta_1^\mathrm{real} - \beta_1^\mathrm{synth}$ \\
\midrule
\multicolumn{3}{c}{\emph{Supervised}} \\
resemblyzer & 0.082 [0.056, 0.116] & 0.56 [0.43, 0.69] \\
ECAPA-TDNN & 0.080 [0.050, 0.118] & 0.09 [0.05, 0.13] \\
WeSpeaker & 0.068 [0.040, 0.104] & 0.10 [0.05, 0.15] \\
TitaNet & 0.085 [0.055, 0.124] & 0.11 [0.07, 0.15] \\
RawNet3 & 0.146 [0.104, 0.195] & 0.15 [0.10, 0.20] \\
x-vector & 0.171 [0.127, 0.230] & 4.14 [3.40, 4.85] \\
\midrule
\multicolumn{3}{c}{\emph{SSL}} \\
WavLM & 0.092 [0.057, 0.144] & 0.63 [0.49, 0.78] \\
wav2vec 2.0 & 0.147 [0.107, 0.198] & 0.73 [0.58, 0.88] \\
HuBERT & 0.140 [0.102, 0.189] & 0.68 [0.53, 0.84] \\
XLS-R & 0.182 [0.131, 0.250] & 2.32 [1.93, 2.73] \\
\midrule
\multicolumn{3}{c}{\emph{Weakly supervised}} \\
Whisper & 0.191 [0.147, 0.318] & 12.00 [1.76, 26.20] \\
\bottomrule
\end{tabular}
\end{table}

\begin{figure}[h]
\centering
\includegraphics[width=0.8\linewidth]{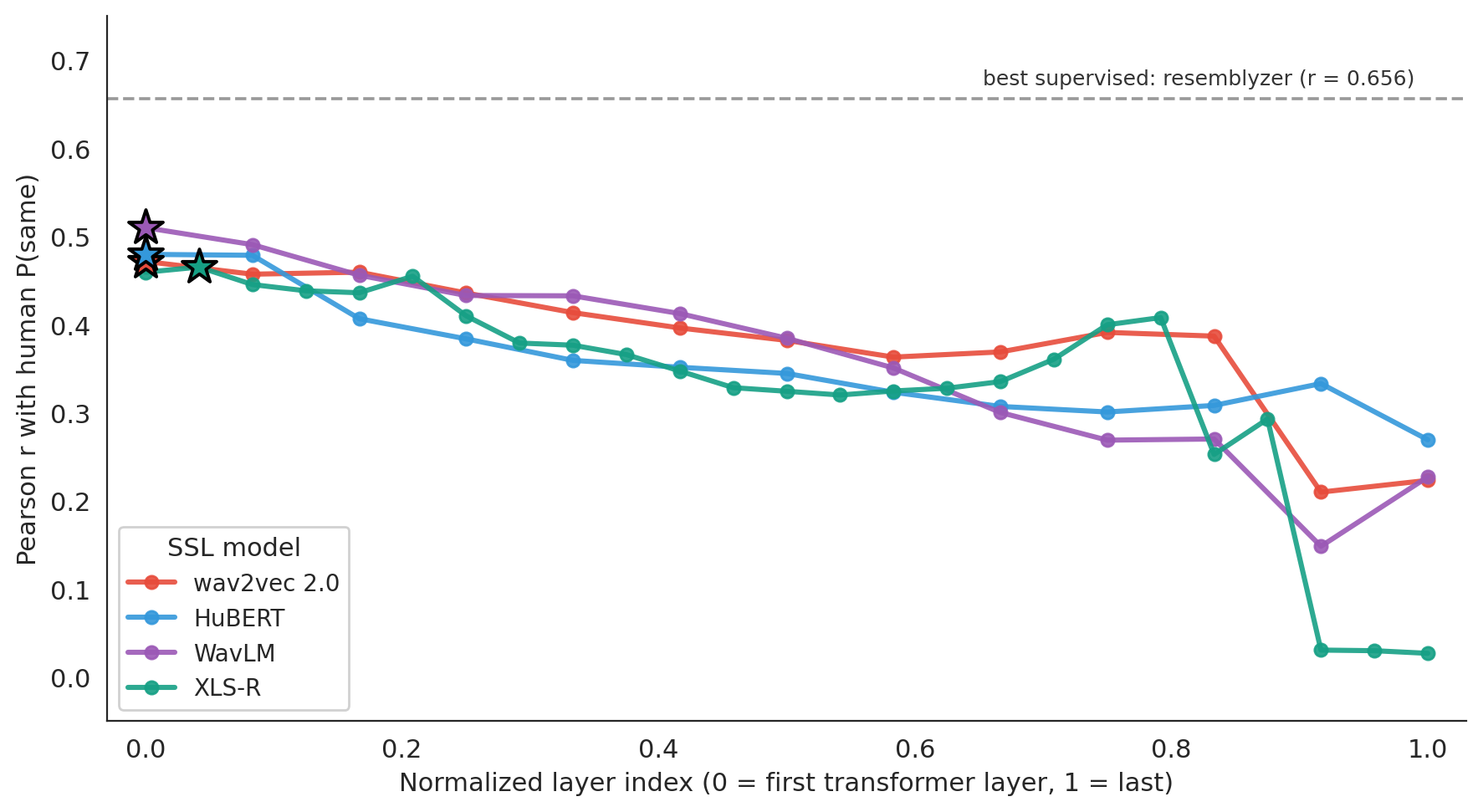}
\caption{Pearson $r$ between the cosine similarity at a layer and $P(\mathrm{same})$ for the four SSL models. The horizontal axis is the layer index divided by the number of layers. Stars mark the layers selected by CV, and the dashed line marks resemblyzer.}
\label{fig:layer-alignment}
\end{figure}

\paragraph{Layers of the SSL models.}
\label{sec:layers-app}
Figure~\ref{fig:layer-alignment} shows Pearson $r$ with $P(\mathrm{same})$ at every layer of the four SSL models, where layer~0 is the input to the first transformer block. For all four models, $r$ is highest at layer~0 or~1 and lower at later layers. XLS-R has $r = 0.47$ at layer~1 and $r = 0.03$ at layer~24. The selected layer is the same in every fold for wav2vec~2.0, WavLM, XLS-R, and Whisper, and HuBERT alternates between layers~0 and~1 (Table~\ref{tab:per-fold-layer}). When we select the layer with the metadata label instead (Table~\ref{tab:metadata-eer}), the selected layer is 0 in eight of the ten folds for wav2vec~2.0 and HuBERT, 4 in every fold for WavLM, and 5 in every fold for XLS-R.

\begin{table}[h]
\centering
\small
\caption{The layer selected in each fold of the CV for the five models with several layers. Column $f_k$ gives the layer that is selected when fold $k$ is held out.}
\label{tab:per-fold-layer}
\begin{tabular}{lcccccccccc}
\toprule
Model & $f_1$ & $f_2$ & $f_3$ & $f_4$ & $f_5$ & $f_6$ & $f_7$ & $f_8$ & $f_9$ & $f_{10}$ \\
\midrule
wav2vec 2.0 & 0 & 0 & 0 & 0 & 0 & 0 & 0 & 0 & 0 & 0 \\
HuBERT      & 1 & 0 & 1 & 0 & 0 & 1 & 0 & 0 & 1 & 0 \\
WavLM       & 0 & 0 & 0 & 0 & 0 & 0 & 0 & 0 & 0 & 0 \\
XLS-R       & 1 & 1 & 1 & 1 & 1 & 1 & 1 & 1 & 1 & 1 \\
Whisper     & 3 & 3 & 3 & 3 & 3 & 3 & 3 & 3 & 3 & 3 \\
\bottomrule
\end{tabular}
\end{table}

\paragraph{Pretrained checkpoints.} Table~\ref{tab:checkpoints} lists the checkpoint and its source for every model in Table~\ref{tab:models}.

\begin{table}[h]
\centering
\small
\caption{Checkpoints and their sources for the eleven models. We load every checkpoint with the loader of its package and do not fine-tune it.}
\label{tab:checkpoints}
\begin{tabular}{lll}
\toprule
Model & Source & Checkpoint identifier \\
\midrule
x-vector     & SpeechBrain         & \texttt{speechbrain/spkrec-xvect-voxceleb} \\
ECAPA-TDNN   & SpeechBrain         & \texttt{speechbrain/spkrec-ecapa-voxceleb} \\
RawNet3      & ESPnet              & \texttt{espnet/voxcelebs12\_rawnet3} \\
TitaNet      & NVIDIA NeMo         & \texttt{nvidia/speakerverification\_en\_titanet\_large} \\
resemblyzer  & \texttt{resemblyzer} (pip) & default voice encoder (GE2E pretrained) \\
WeSpeaker    & pyannote.audio 3.4  & \texttt{pyannote/wespeaker-voxceleb-resnet34-LM} \\
wav2vec 2.0  & Hugging Face        & \texttt{facebook/wav2vec2-base} \\
HuBERT       & Hugging Face        & \texttt{facebook/hubert-base-ls960} \\
WavLM        & Hugging Face        & \texttt{microsoft/wavlm-base-plus} \\
XLS-R        & Hugging Face        & \texttt{facebook/wav2vec2-xls-r-300m} \\
Whisper      & OpenAI \texttt{whisper} (pip) & \texttt{base} (74M params, 512-dim encoder) \\
\bottomrule
\end{tabular}
\end{table}

\paragraph{Models on the metadata target.}
\label{sec:metadata-results}
We evaluate the eleven models against the metadata label of each pair on three sets of pairs (Table~\ref{tab:metadata-eer}). Set~A has the 800 pairs with a real comparison recording of the same speaker or of a different speaker, and it is the set closest to a speaker verification benchmark. On Set~A, four supervised models have an equal error rate (EER) from 1.5\% to 1.9\% (WeSpeaker, TitaNet, ECAPA-TDNN, and RawNet3). x-vector and resemblyzer have 11.1\% and 12.0\%, the SSL models have 26\% to 32\%, and Whisper has 37.5\%. Set~B has the 800 pairs with a voice clone. The Set~B EERs are 3.0\% to 4.6\% for the four best models, 15.2\% and 16.0\% for resemblyzer and x-vector, 32\% to 37\% for the SSL models, and 43.1\% for Whisper. The EER of every model is higher on Set~B than on Set~A, by 1.4 to 7.3 percentage points, and the rank correlation between the two orderings is 0.97. Set~C has the 1{,}700 pairs that are not morphs. The metadata target and the perception target order the three types of models in the same way, and they order the supervised models differently. Resemblyzer has the highest $r$ in Task~1 and the highest EER of the six supervised models on Set~A. The rank correlation between the metadata AUC on Set~C and the $r$ with $P(\mathrm{same})$ on the same pairs is 0.84.

\begin{table}[h]
\centering
\small
\caption{Metadata target on three sets of pairs. Set~A has the pairs with a real comparison recording of the same speaker or of a different speaker (800 pairs). Set~B has the pairs with a voice clone (800 pairs). Set~C has all pairs that are not morphs (1{,}700 pairs). For the SSL models and Whisper, the layer is selected with the metadata label by the CV over speakers. Models are sorted by the EER on Set~A within each type.}
\label{tab:metadata-eer}
\begin{tabular}{lrrrrrr}
\toprule
 & \multicolumn{3}{c}{EER} & \multicolumn{3}{c}{AUC} \\
\cmidrule(lr){2-4} \cmidrule(lr){5-7}
Model & Set A & Set B & Set C & Set A & Set B & Set C \\
\midrule
\multicolumn{7}{c}{\emph{Supervised}} \\
WeSpeaker & \textbf{1.5\%} & 3.5\% & 3.6\% & 0.998 & 0.994 & 0.995 \\
TitaNet & 1.6\% & \textbf{3.0\%} & \textbf{2.7\%} & \textbf{0.999} & \textbf{0.995} & \textbf{0.997} \\
ECAPA-TDNN & 1.6\% & 4.6\% & 3.1\% & 0.998 & 0.992 & 0.995 \\
RawNet3 & 1.9\% & 4.0\% & 3.5\% & 0.997 & 0.992 & 0.995 \\
x-vector & 11.1\% & 16.0\% & 11.9\% & 0.955 & 0.937 & 0.952 \\
resemblyzer & 12.0\% & 15.2\% & 13.0\% & 0.953 & 0.934 & 0.949 \\
\midrule
\multicolumn{7}{c}{\emph{SSL}} \\
XLS-R & 26.4\% & 32.0\% & 27.8\% & 0.795 & 0.726 & 0.785 \\
WavLM & 26.5\% & 33.8\% & 28.2\% & 0.805 & 0.716 & 0.781 \\
HuBERT & 31.1\% & 34.8\% & 31.2\% & 0.747 & 0.715 & 0.760 \\
wav2vec 2.0 & 32.0\% & 37.0\% & 31.8\% & 0.734 & 0.694 & 0.745 \\
\midrule
\multicolumn{7}{c}{\emph{Weakly supervised}} \\
Whisper & 37.5\% & 43.1\% & 40.2\% & 0.673 & 0.603 & 0.654 \\
\bottomrule
\end{tabular}
\end{table}

\begin{figure}[h]
\centering
\includegraphics[width=0.8\linewidth]{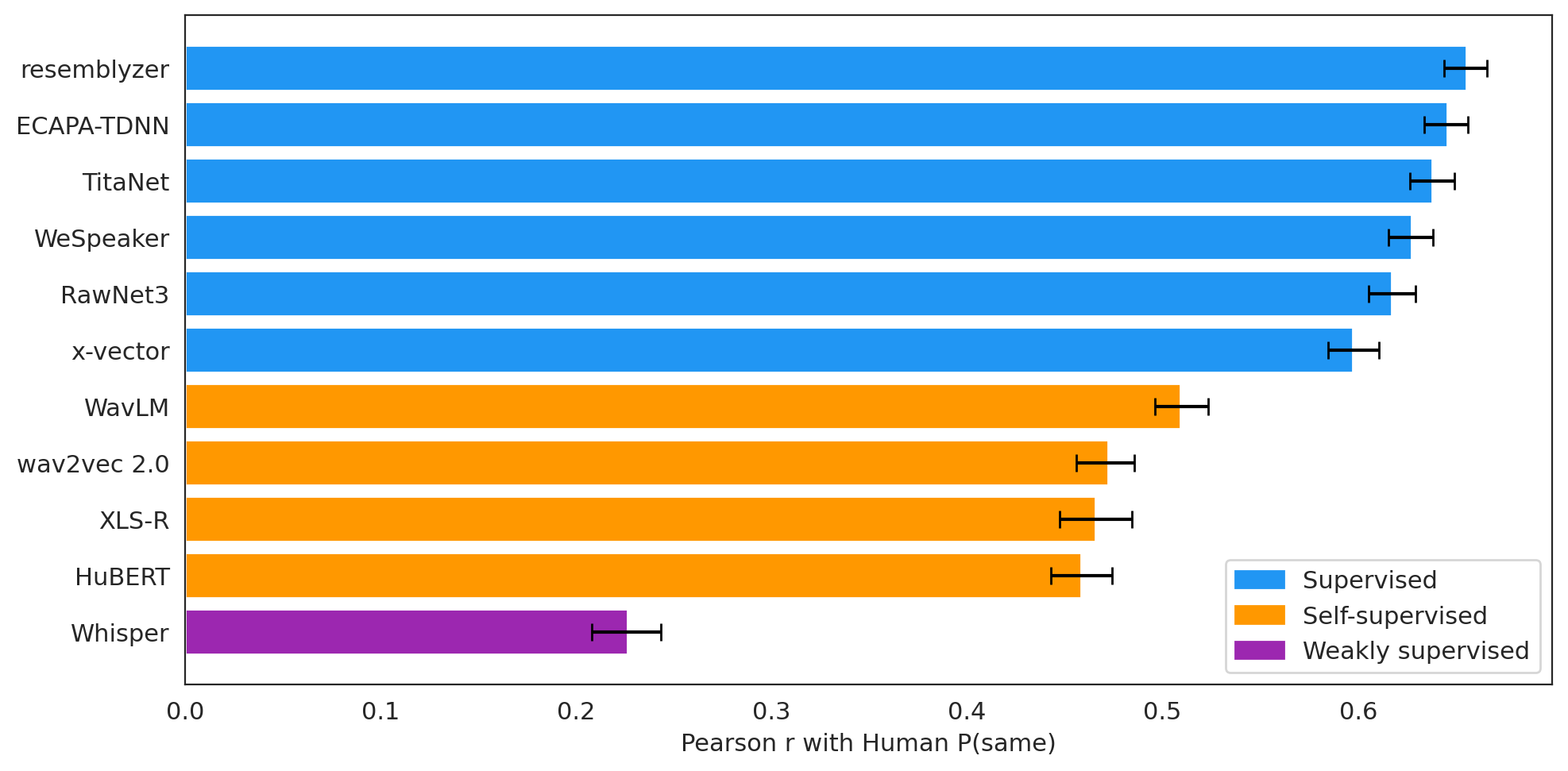}
\caption{Pearson $r$ between the cosine similarity of each model and $P(\mathrm{same})$ on all pairs, with 95\% bootstrap confidence intervals (1{,}000 resamples of the pairs). The SSL models and Whisper use the layer selected by CV.}
\label{fig:pearson-bar}
\end{figure}

\begin{figure}[h]
\centering
\includegraphics[width=0.7\linewidth]{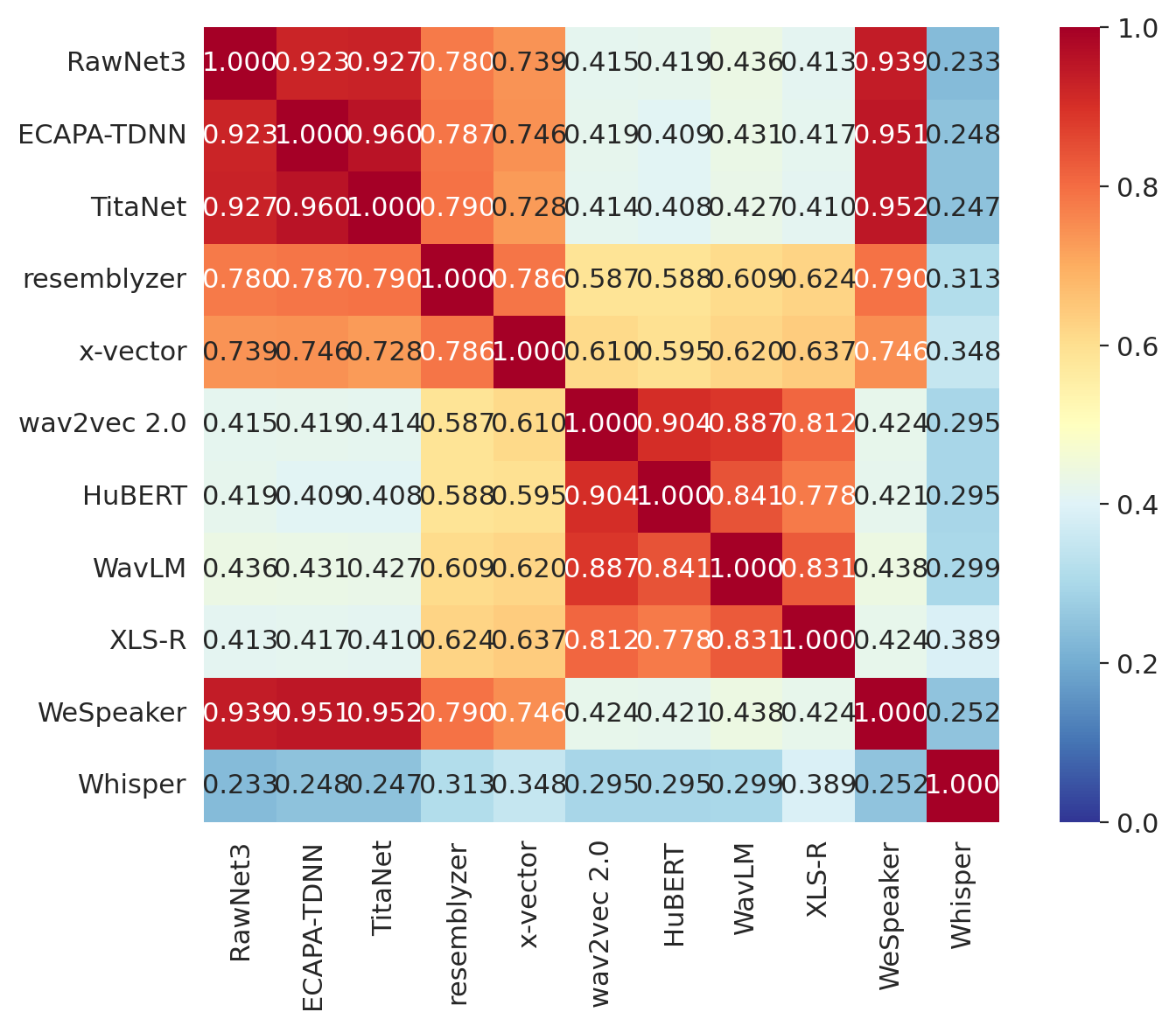}
\caption{Pearson correlation between the cosine similarities of every two models over the 9{,}800 pairs. Warmer colors mark higher correlations. The correlation is 0.73 to 0.96 between two supervised models, 0.78 to 0.90 between two SSL models, and 0.41 to 0.64 between a supervised model and an SSL model. Whisper has a correlation of 0.23 to 0.39 with the other models.}
\label{fig:intermodel}
\end{figure}

\begin{figure}[h]
\centering
\includegraphics[width=0.85\linewidth]{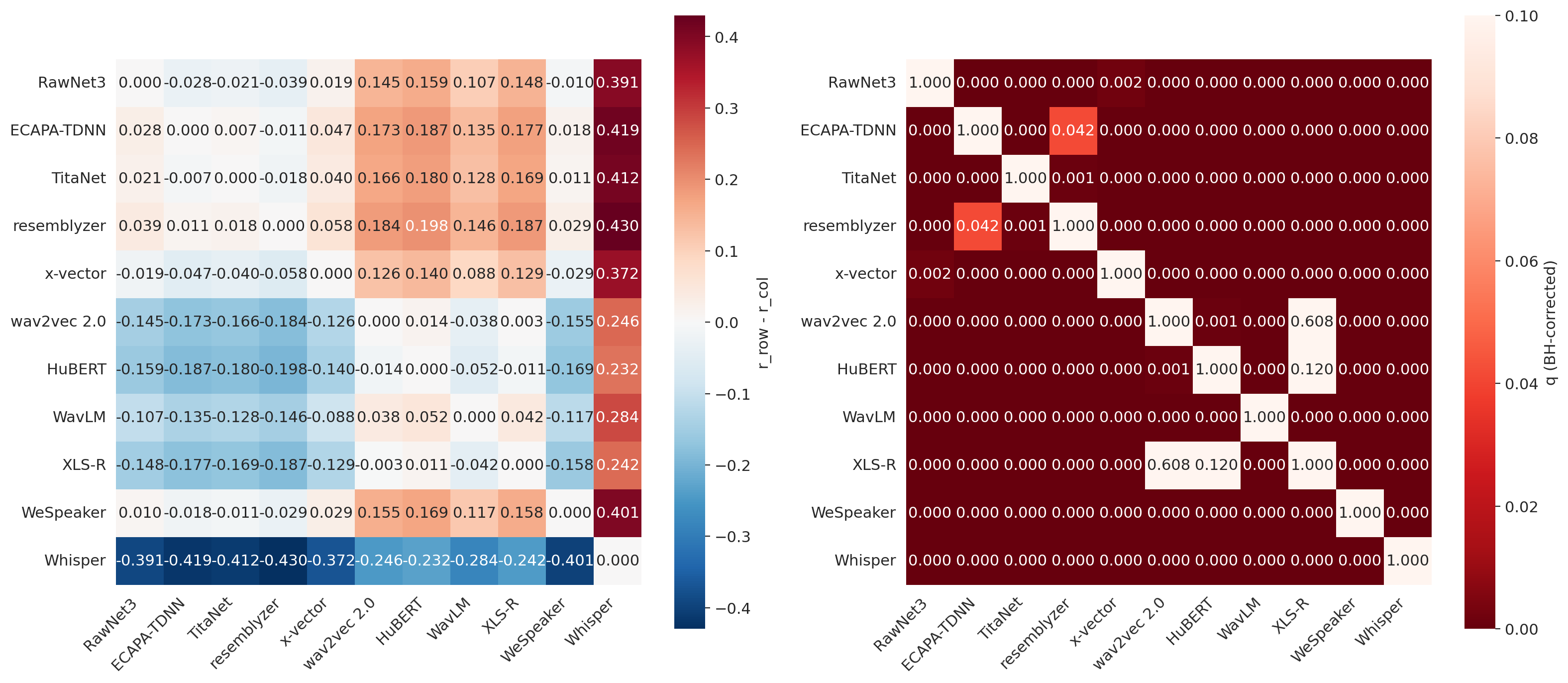}
\caption{Comparison of every two models in Task~1 with the pairs resampled 2{,}000 times, on the 9{,}799 pairs that every model can score, with Benjamini-Hochberg correction across the 55 comparisons. \textbf{Left.} Mean difference $r_\mathrm{row} - r_\mathrm{col}$ over the resamples, which is red when the row model has the higher $r$. \textbf{Right.} Corrected $p$-values ($q$), where darker cells have smaller values. Of the 55 comparisons, 53 are significant at 0.05. The two that are not significant are wav2vec~2.0 with XLS-R ($q = 0.61$) and HuBERT with XLS-R ($q = 0.12$).}
\label{fig:pairwise-sig}
\end{figure}

\paragraph{Task~1 figures.} Figure~\ref{fig:pearson-bar} shows $r$ for each model with 95\% bootstrap confidence intervals. Figure~\ref{fig:intermodel} shows the correlation between the cosine similarities of every two models. Figure~\ref{fig:pairwise-sig} shows the comparison of every two models with the pairs resampled.

\begin{figure}[h]
\centering
\includegraphics[width=0.8\linewidth]{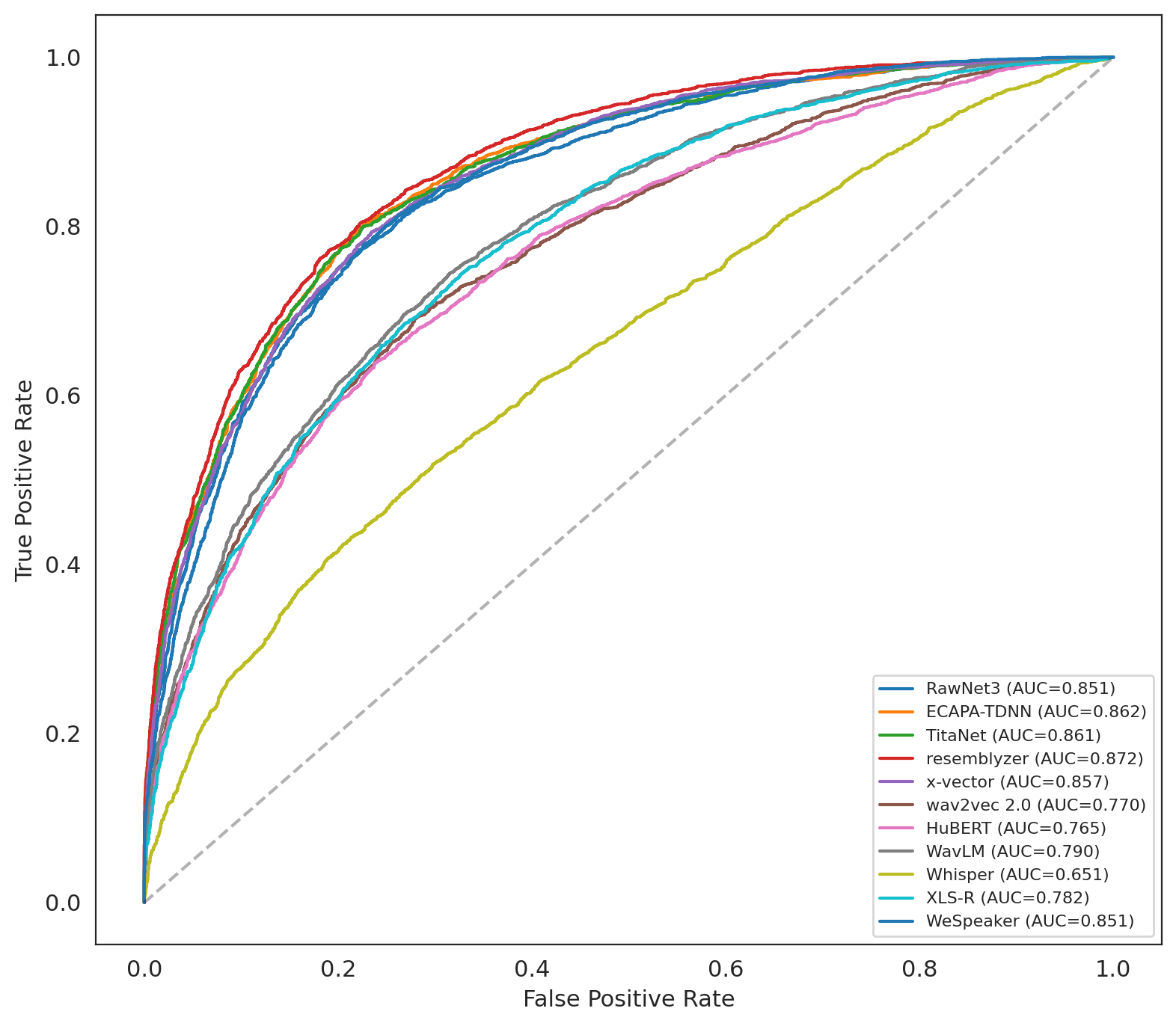}
\caption{ROC curves of Task~2, with the majority answer of the listeners as the label ($P(\mathrm{same}) > 0.5$, on the 8{,}884 pairs without a tied vote). Each curve is one model. The supervised models reach an AUC of 0.85 to 0.87, the SSL models 0.77 to 0.79, and Whisper 0.65. The dashed diagonal marks chance.}
\label{fig:roc-curves}
\end{figure}

\paragraph{Task~2 ROC curves.} Figure~\ref{fig:roc-curves} shows the ROC curve of each model for the majority label.

\paragraph{Individual human baseline.}
\label{sec:human-baseline-app}
For each of the 806 listeners with at least 10 trials, we compute the share of the listener's trials in which the answer agrees with the majority of the other listeners on the pair. Pairs with a tied vote of the other listeners are left out. We require 10 trials because the agreement of a listener with fewer trials can take only a few values. The mean agreement is 0.73 (median 0.74, SD 0.12, range 0.24 to 1.00). For a model, we choose the cutoff on the cosine similarity that maximizes Youden's $J$ against the majority label, and the accuracy of the model is its agreement with the majority of all listeners on the 8{,}884 pairs without a tied vote. A listener is therefore scored on the pairs that the listener judged, against the other listeners, and a model is scored on all pairs, against all listeners.

\begin{figure}[h]
\centering
\includegraphics[width=0.85\linewidth]{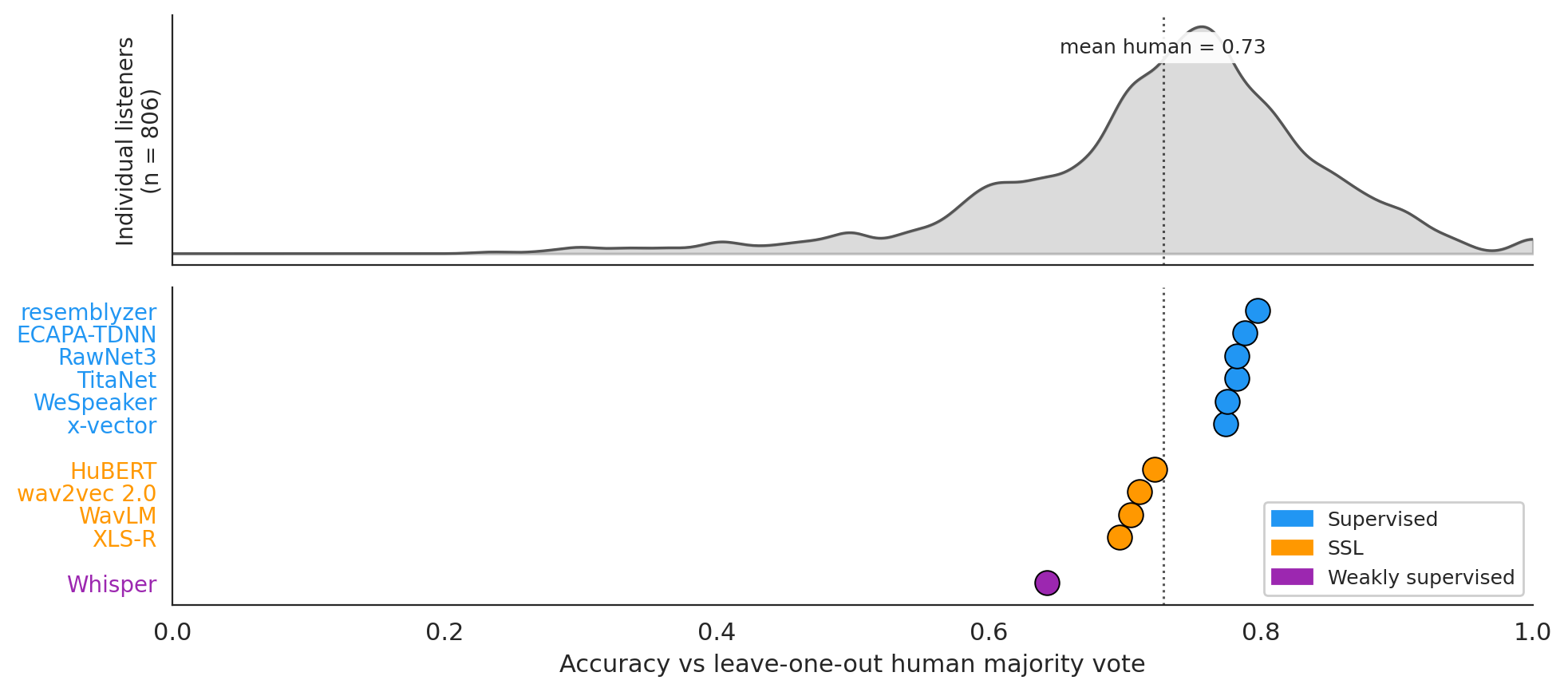}
\caption{\textbf{Top.} Distribution of the agreement of the 806 listeners with the majority of the other listeners. \textbf{Bottom.} Accuracy of each model against the majority of all listeners, with the cutoff chosen on all pairs. The dotted line marks the mean agreement of the listeners (0.73).}
\label{fig:human-baseline}
\end{figure}

Table~\ref{tab:nested-thresholds} reports the accuracy of each model and the share of listeners below it for two cutoffs. The first cutoff is chosen on the pairs of the nine training folds and applied to the held-out fold, and the main text reports it. The second is chosen on all pairs, as in Figure~\ref{fig:human-baseline}. The accuracy differs by at most 0.007 between the two cutoffs, and the share of listeners by at most 2.9 percentage points. The order of the models is the same, except that WeSpeaker and x-vector exchange places, and so do WavLM and XLS-R, with differences in accuracy below 0.005.

For each model, we also correlated the distance of the cosine similarity from the cutoff with the agreement among the listeners on the pair, which we measure as 1 minus the entropy of the votes. The correlation is 0.34 for resemblyzer, 0.28 to 0.30 for the other supervised models, 0.21 to 0.24 for the SSL models, and 0.11 for Whisper.

\begin{table}[h]
\centering
\small
\caption{Accuracy of each model against the majority of all listeners, and the share of the 806 listeners whose agreement is below it. The cutoff on the cosine similarity is chosen with the CV over speakers or on all pairs.}
\label{tab:nested-thresholds}
\begin{tabular}{lcccc}
\toprule
 & \multicolumn{2}{c}{Cutoff from the CV} & \multicolumn{2}{c}{Cutoff from all pairs} \\
\cmidrule(lr){2-3} \cmidrule(lr){4-5}
Model & Accuracy & Listeners below & Accuracy & Listeners below \\
\midrule
\multicolumn{5}{c}{\emph{Supervised}} \\
resemblyzer & 0.790 & 72.5\% & 0.797 & 74.3\% \\
ECAPA-TDNN & 0.784 & 70.1\% & 0.788 & 71.7\% \\
TitaNet & 0.782 & 69.4\% & 0.782 & 69.6\% \\
RawNet3 & 0.778 & 68.9\% & 0.782 & 69.6\% \\
x-vector & 0.774 & 66.7\% & 0.774 & 66.9\% \\
WeSpeaker & 0.770 & 64.6\% & 0.775 & 67.0\% \\
\midrule
\multicolumn{5}{c}{\emph{SSL}} \\
HuBERT & 0.718 & 38.7\% & 0.722 & 40.2\% \\
wav2vec 2.0 & 0.710 & 36.0\% & 0.711 & 36.5\% \\
XLS-R & 0.703 & 33.0\% & 0.696 & 30.1\% \\
WavLM & 0.703 & 33.0\% & 0.704 & 33.0\% \\
\midrule
\multicolumn{5}{c}{\emph{Weakly supervised}} \\
Whisper & 0.646 & 19.5\% & 0.643 & 19.1\% \\
\bottomrule
\end{tabular}
\end{table}

\paragraph{Consistency of the listeners.}
\label{sec:consistency-app}
No listener judged the same pair twice, so the judgments give no test-retest agreement. The morphs give a close substitute. For each reference speaker and comparison speaker, two morphs with the same weight are made from different source recordings. In 6{,}332 cases, one listener judged both morphs (118 listeners), and the two answers agreed in 66\% of the cases. Two different listeners would agree in 58\% of the cases, given the $P(\mathrm{same})$ of the two morphs. The 95\% interval of the difference is 0.01 to 0.19, from 2{,}000 resamples of the listeners. The listeners also differ from each other in how often they answer ``same''. For each listener, we compute the mean difference between the listener's answer and the $P(\mathrm{same})$ of the pair, separately in two random halves of the listener's trials. The two values have a correlation of 0.72 across the 393 listeners with at least 30 trials. On the pairs from one recording, the median listener answers ``different'' in 4.5\% of the trials (126 listeners with at least 5 such trials).

\begin{figure}[h]
\centering
\includegraphics[width=0.95\linewidth]{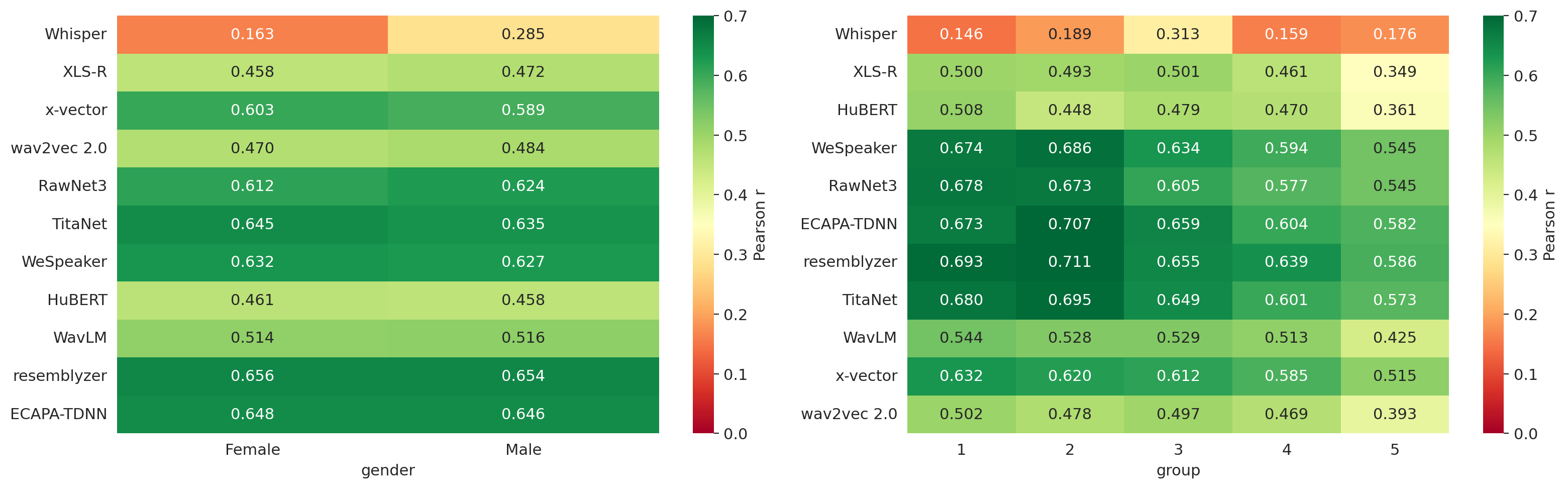}
\caption{Pearson $r$ between the cosine similarity and $P(\mathrm{same})$ on the pairs of each gender (left) and of each sociophonetic group (right). The difference between female and male speakers is at most 0.014 for every model except Whisper (0.12). The difference between the highest and the lowest group is 0.11 to 0.17 for every model.}
\label{fig:fairness}
\end{figure}

\paragraph{Demographic fairness.} Figure~\ref{fig:fairness} shows the $r$ of Task~1 on the pairs of each gender and of each sociophonetic group (Section~\ref{sec:fairness}).

\paragraph{Ranking with the reference speakers resampled.}
\label{sec:speaker-resampling-app}
The comparisons in Task~1 resample pairs, and the 98 pairs of a reference speaker share its reference recording. We therefore also resampled the 100 reference speakers with all their pairs 1{,}000 times (Table~\ref{tab:speaker-resampling}). Every supervised model is above every SSL model, and every SSL model is above Whisper, in all resamples. Resemblyzer has the highest $r$ in 81.0\% of the resamples, ECAPA-TDNN in 18.4\%, and TitaNet in 0.6\%. With speakers resampled, 46 of the 55 comparisons between models are significant. The nine that are not significant are the three among resemblyzer, ECAPA-TDNN, and TitaNet, WeSpeaker with TitaNet and with RawNet3, RawNet3 with x-vector, and the three among wav2vec~2.0, XLS-R, and HuBERT.

\begin{table}[h]
\centering
\small
\caption{Task~1 with the 100 reference speakers resampled 1{,}000 times. The rank is the position of a model among the eleven models in a resample.}
\label{tab:speaker-resampling}
\begin{tabular}{lcccc}
\toprule
Model & $r$ & 95\% interval of $r$ & Mean rank & 95\% interval of the rank \\
\midrule
\multicolumn{5}{c}{\emph{Supervised}} \\
resemblyzer & 0.656 & [0.632, 0.679] & 1.26 & 1 to 3 \\
ECAPA-TDNN & 0.646 & [0.619, 0.672] & 1.87 & 1 to 3 \\
TitaNet & 0.638 & [0.611, 0.664] & 2.91 & 2 to 4 \\
WeSpeaker & 0.627 & [0.600, 0.653] & 4.06 & 3 to 5 \\
RawNet3 & 0.617 & [0.586, 0.645] & 4.96 & 4 to 6 \\
x-vector & 0.597 & [0.572, 0.623] & 5.94 & 5 to 6 \\
\midrule
\multicolumn{5}{c}{\emph{SSL}} \\
WavLM & 0.510 & [0.474, 0.543] & 7.00 & 7 to 7 \\
wav2vec 2.0 & 0.471 & [0.432, 0.508] & 8.38 & 8 to 10 \\
XLS-R & 0.465 & [0.432, 0.497] & 8.99 & 8 to 10 \\
HuBERT & 0.458 & [0.418, 0.496] & 9.63 & 8 to 10 \\
\midrule
\multicolumn{5}{c}{\emph{Weakly supervised}} \\
Whisper & 0.226 & [0.177, 0.268] & 11.00 & 11 to 11 \\
\bottomrule
\end{tabular}
\end{table}

\paragraph{Robustness of the Task~1 ranking.}
\label{sec:robustness-app}
Table~\ref{tab:robustness} reports $r$ under four changes of the data. On the 900 pairs of real recordings alone, the rank correlation with the full ranking is 0.98. Without the 8{,}100 morphs it is also 0.98, and $r$ is higher by 0.10 to 0.16 for every model except Whisper. With equal weight for each of the six stimulus types, which is the mean of the six values of $r$ in Table~\ref{tab:by-type-r}, the rank correlation is 0.90, and RawNet3 falls below WavLM and XLS-R. RawNet3 has a low $r$ on the pairs from one recording (0.18), whose $P(\mathrm{same})$ values span a narrow range. Without that type the rank correlation is 0.92, and every supervised model is above every SSL model. The types with a real comparison recording have 37 to 40 judgments for each pair, and the other types have about 10. When we reduce every pair of the former types to 10 judgments, the ranking is identical to the full ranking in each of 20 random draws.

\begin{table}[h]
\centering
\small
\caption{Task~1 $r$ on all pairs, on the 900 pairs of real recordings, on the 1{,}700 pairs that are not morphs, with equal weight for each of the six stimulus types, and with 10 judgments for each pair (mean of 20 draws).}
\label{tab:robustness}
\begin{tabular}{lccccc}
\toprule
Model & All pairs & Real recordings & Without morphs & Equal weight & 10 judgments \\
\midrule
\multicolumn{6}{c}{\emph{Supervised}} \\
resemblyzer & 0.656 & 0.859 & 0.804 & 0.506 & 0.650 \\
ECAPA-TDNN & 0.646 & 0.855 & 0.791 & 0.442 & 0.640 \\
TitaNet & 0.638 & 0.858 & 0.794 & 0.444 & 0.633 \\
WeSpeaker & 0.627 & 0.847 & 0.784 & 0.430 & 0.622 \\
RawNet3 & 0.617 & 0.843 & 0.774 & 0.381 & 0.612 \\
x-vector & 0.597 & 0.830 & 0.762 & 0.439 & 0.592 \\
\midrule
\multicolumn{6}{c}{\emph{SSL}} \\
WavLM & 0.510 & 0.673 & 0.615 & 0.394 & 0.505 \\
wav2vec 2.0 & 0.471 & 0.639 & 0.573 & 0.352 & 0.468 \\
XLS-R & 0.465 & 0.653 & 0.591 & 0.390 & 0.461 \\
HuBERT & 0.458 & 0.626 & 0.568 & 0.339 & 0.454 \\
\midrule
\multicolumn{6}{c}{\emph{Weakly supervised}} \\
Whisper & 0.226 & 0.213 & 0.203 & 0.057 & 0.224 \\
\bottomrule
\end{tabular}
\end{table}

\begin{table}[h]
\centering
\small
\caption{Task~1 $r$ inside each stimulus type. The last row gives the reliability of $P(\mathrm{same})$ inside the type (Table~\ref{tab:type-stats}).}
\label{tab:by-type-r}
\begin{tabular}{lcccccc}
\toprule
 & \multicolumn{3}{c}{Same speaker} & \multicolumn{2}{c}{Different speaker} & \\
\cmidrule(lr){2-4} \cmidrule(lr){5-6}
Model & One recording & Two recordings & Clone & Real & Clone & Morph \\
\midrule
\multicolumn{7}{c}{\emph{Supervised}} \\
resemblyzer & 0.286 & 0.650 & 0.513 & 0.520 & 0.463 & 0.607 \\
ECAPA-TDNN & 0.348 & 0.581 & 0.453 & 0.365 & 0.306 & 0.598 \\
TitaNet & 0.246 & 0.612 & 0.498 & 0.409 & 0.316 & 0.585 \\
WeSpeaker & 0.341 & 0.550 & 0.407 & 0.375 & 0.332 & 0.572 \\
RawNet3 & 0.180 & 0.497 & 0.336 & 0.386 & 0.314 & 0.571 \\
x-vector & 0.232 & 0.558 & 0.451 & 0.472 & 0.370 & 0.550 \\
\midrule
\multicolumn{7}{c}{\emph{SSL}} \\
WavLM & 0.318 & 0.453 & 0.413 & 0.392 & 0.318 & 0.467 \\
wav2vec 2.0 & 0.247 & 0.403 & 0.360 & 0.364 & 0.305 & 0.435 \\
XLS-R & 0.304 & 0.510 & 0.414 & 0.361 & 0.325 & 0.426 \\
HuBERT & 0.173 & 0.414 & 0.366 & 0.342 & 0.320 & 0.417 \\
\midrule
\multicolumn{7}{c}{\emph{Weakly supervised}} \\
Whisper & $-0.084$ & $-0.007$ & 0.064 & 0.063 & 0.090 & 0.216 \\
\midrule
Reliability & 0.59 & 0.87 & 0.58 & 0.85 & 0.52 & 0.66 \\
\bottomrule
\end{tabular}
\end{table}

\paragraph{Pairs outside VoxCeleb.}
\label{sec:voxceleb-app}
We matched the names of the 100 speakers against the speaker lists of VoxCeleb~1 and VoxCeleb~2, and 26 speakers match (11 in VoxCeleb~1 and 15 in VoxCeleb~2). Of the 9{,}800 pairs, 5{,}580 have both speakers outside VoxCeleb. Table~\ref{tab:voxceleb} reports $r$ on these pairs and on the other 4{,}220 pairs. The rank correlation between the ranking on the 5{,}580 pairs and the full ranking is 0.96, and the order of the three types of models is the same. ECAPA-TDNN has the highest $r$ on the 5{,}580 pairs. The mean $r$ of the supervised models is 0.166 above the mean $r$ of the SSL models on the 5{,}580 pairs and 0.142 above it on the other pairs.

\begin{table}[h]
\centering
\small
\caption{Task~1 $r$ on all pairs, on the 5{,}580 pairs with both speakers outside VoxCeleb, and on the 4{,}220 pairs with at least one speaker in VoxCeleb.}
\label{tab:voxceleb}
\begin{tabular}{lccc}
\toprule
Model & All pairs & Outside VoxCeleb & With a VoxCeleb speaker \\
\midrule
\multicolumn{4}{c}{\emph{Supervised}} \\
resemblyzer & 0.656 & 0.664 & 0.642 \\
ECAPA-TDNN & 0.646 & 0.672 & 0.608 \\
TitaNet & 0.638 & 0.671 & 0.589 \\
WeSpeaker & 0.627 & 0.650 & 0.592 \\
RawNet3 & 0.617 & 0.635 & 0.592 \\
x-vector & 0.597 & 0.619 & 0.574 \\
\midrule
\multicolumn{4}{c}{\emph{SSL}} \\
WavLM & 0.510 & 0.515 & 0.494 \\
wav2vec 2.0 & 0.471 & 0.486 & 0.443 \\
XLS-R & 0.465 & 0.464 & 0.459 \\
HuBERT & 0.458 & 0.477 & 0.432 \\
\midrule
\multicolumn{4}{c}{\emph{Weakly supervised}} \\
Whisper & 0.226 & 0.252 & 0.192 \\
\bottomrule
\end{tabular}
\end{table}

\begin{figure}[h]
\centering
\includegraphics[width=0.75\linewidth]{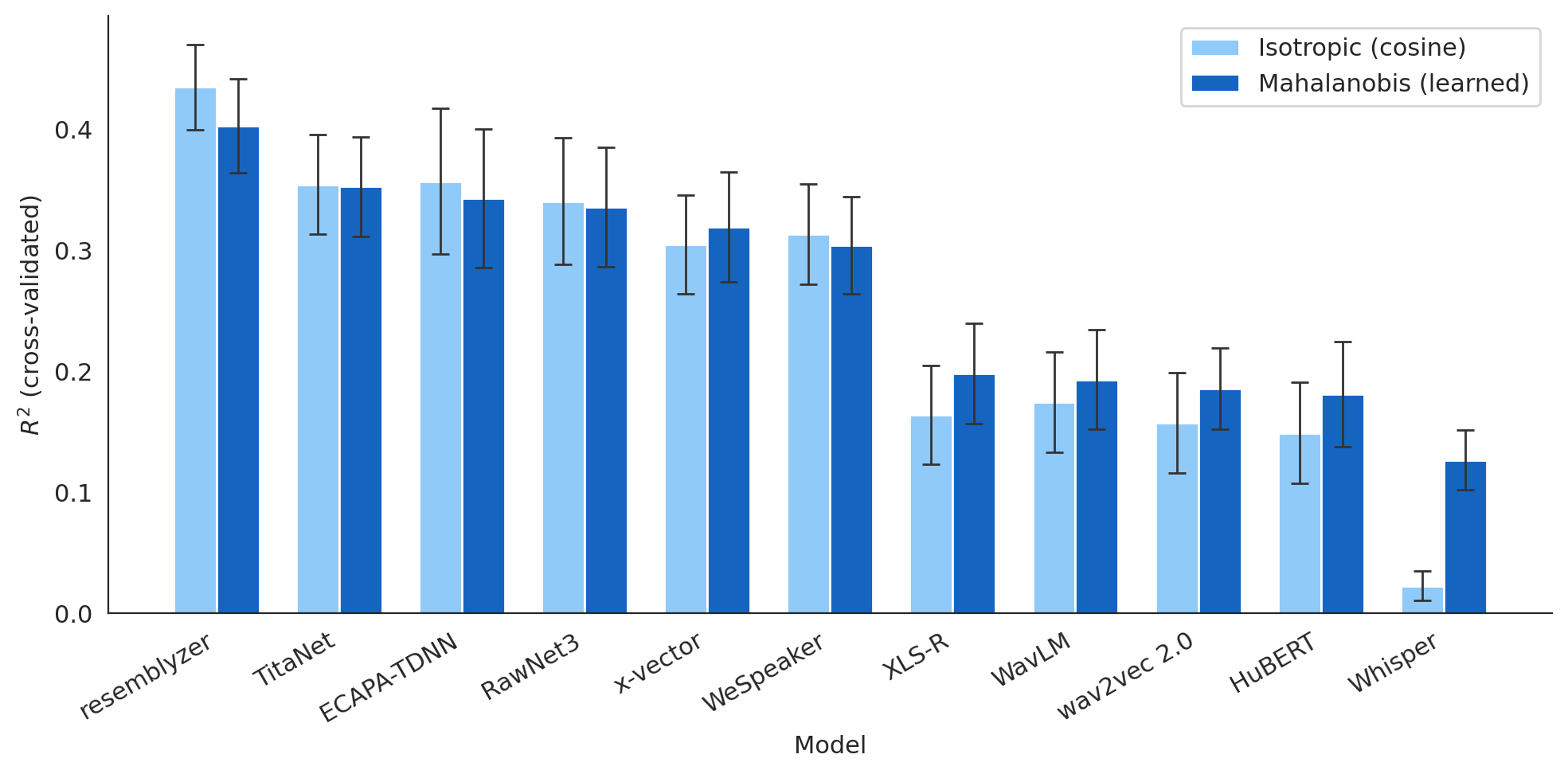}
\caption{$R^2$ with $P(\mathrm{same})$ with equal weights (light blue) and with the learned weights of the diagonal Mahalanobis metric (dark blue), with the CV over speakers. Error bars are 95\% confidence intervals across the 10 folds.}
\label{fig:mahalanobis}
\end{figure}

\paragraph{Learned similarity functions.}
\label{sec:learned-app}
\label{sec:fairness-isotropy}
We fitted three learned similarity functions with the CV over speakers. The first is a diagonal Mahalanobis metric, which learns one weight for each dimension of the embedding (Table~\ref{tab:mahalanobis}, Figure~\ref{fig:mahalanobis}). For the models with more than 256 dimensions, we first reduce the embeddings to 50 principal components. The table compares the $R^2$ with learned weights with the $R^2$ with equal weights on the same features. The learned weights change $R^2$ by $-0.03$ to $+0.01$ for the supervised models, by $+0.02$ to $+0.03$ for the SSL models, and by $+0.10$ for Whisper. The second function is a ridge regression on all products of two of the first 50 principal components of the difference between the two normalized embeddings (1{,}275 features). The third is a ridge regression on the features $[\,|\mathbf{e}_\mathrm{ref} - \mathbf{e}_\mathrm{comp}|,\ \mathbf{e}_\mathrm{ref} \odot \mathbf{e}_\mathrm{comp}\,]$ of the normalized embeddings. Table~\ref{tab:probes} reports the $r$ between the predictions for the held-out folds and $P(\mathrm{same})$. With the third function, the supervised models have an $r$ from 0.60 to 0.68, and the SSL models from 0.57 to 0.58. The difference between the weakest supervised model and the strongest SSL model is 0.09 with the cosine similarity, 0.04 with the second function, and 0.02 with the third. Every supervised model is above every SSL model, and every SSL model is above Whisper, for all three functions.

\begin{table}[h]
\centering
\small
\caption{Diagonal Mahalanobis metric with the CV over speakers. $R^2_\mathrm{iso}$ uses equal weights and $R^2_\mathrm{mahal}$ uses learned weights on the same features, and $\Delta$ is their difference. For x-vector, the SSL models, and Whisper, the features are 50 principal components, so $R^2_\mathrm{iso}$ differs from the $R^2$ in Table~\ref{tab:grand-summary}.}
\label{tab:mahalanobis}
\begin{tabular}{lccc}
\toprule
Model & $R^2_\mathrm{iso}$ & $R^2_\mathrm{mahal}$ & $\Delta$ \\
\midrule
\multicolumn{4}{c}{\emph{Supervised}} \\
resemblyzer & 0.435 & 0.403 & $-0.032$ \\
ECAPA-TDNN & 0.357 & 0.343 & $-0.014$ \\
TitaNet & 0.354 & 0.353 & $-0.002$ \\
RawNet3 & 0.341 & 0.336 & $-0.005$ \\
WeSpeaker & 0.314 & 0.304 & $-0.009$ \\
x-vector & 0.305 & 0.319 & $+0.014$ \\
\midrule
\multicolumn{4}{c}{\emph{SSL}} \\
WavLM & 0.175 & 0.193 & $+0.019$ \\
XLS-R & 0.164 & 0.198 & $+0.034$ \\
wav2vec 2.0 & 0.157 & 0.186 & $+0.028$ \\
HuBERT & 0.149 & 0.181 & $+0.032$ \\
\midrule
\multicolumn{4}{c}{\emph{Weakly supervised}} \\
Whisper & 0.023 & 0.127 & $+0.104$ \\
\bottomrule
\end{tabular}
\end{table}

\begin{table}[h]
\centering
\small
\caption{Task~1 $r$ with the cosine similarity, with the ridge regression on products of principal components of the embedding difference (quadratic), and with the ridge regression on pair features.}
\label{tab:probes}
\begin{tabular}{lccc}
\toprule
Model & Cosine & Quadratic & Pair features \\
\midrule
\multicolumn{4}{c}{\emph{Supervised}} \\
resemblyzer & 0.656 & 0.619 & 0.680 \\
ECAPA-TDNN & 0.646 & 0.559 & 0.624 \\
TitaNet & 0.638 & 0.574 & 0.601 \\
WeSpeaker & 0.627 & 0.570 & 0.614 \\
RawNet3 & 0.617 & 0.575 & 0.605 \\
x-vector & 0.597 & 0.576 & 0.642 \\
\midrule
\multicolumn{4}{c}{\emph{SSL}} \\
WavLM & 0.510 & 0.523 & 0.574 \\
wav2vec 2.0 & 0.471 & 0.488 & 0.567 \\
XLS-R & 0.465 & 0.504 & 0.570 \\
HuBERT & 0.458 & 0.515 & 0.584 \\
\midrule
\multicolumn{4}{c}{\emph{Weakly supervised}} \\
Whisper & 0.226 & 0.294 & 0.423 \\
\bottomrule
\end{tabular}
\end{table}

\FloatBarrier

\end{document}